# Mesons and diquarks in a color superconducting regime

Eric Blanquier
*12 Boulevard des capucines, F-11 800 Trèbes, France*
*E-mail: ericblanquier@hotmail.com*

**Abstract**
The behavior of the mesons and diquarks is studied at finite temperatures, chemical potentials and densities, notably when the color superconductivity is taken into account. The Nambu and Jona-Lasinio model complemented by a Polyakov loop (PNJL description) has been adapted in order to model these composite particles in this regime. This paper focuses on the scalar and pseudo-scalar mesons and diquarks, in a three-flavor and three-color description, with the isospin symmetry and at null strangeness. A finality of this work is to underline the modifications carried out by the color superconducting regime on the used equations and on the obtained results. It has been observed that the two-flavor color-superconducting (2SC) phase affects the masses of the mesons and diquarks in a non-negligible way. This observation is particularly true at high densities and low temperatures for the pions, $\eta$ and the diquarks $[ud]$ whose color is $rg$. This reveals that the inclusion of the color superconductivity is relevant to describe the mesons and diquarks near the first order chiral phase transition.

## I. INTRODUCTION

After the exciting results supplied by the RHIC and LHC concerning the quark gluon plasma [1,2], the next step may concern the exploration of low temperatures and high densities zones of the phase diagram. Future accelerators like the FAIR and NICA are expected to reach this objective [3,4]. In this regime, the conditions are favorable to the formation of quark-quark pairs. They are the Cooper pairs of the color superconductivity [5-8]. This phenomenon is expected to play a great role in the evolution of compact stars, including the neutron stars, quark stars, strange stars or magnetars [9-13]. In the literature, it has been imagined various pairings between the quarks. They correspond to several color superconducting phases [14]. The most famous ones are the 2-flavor color-superconducting (2SC) [15] and color-flavor–locked (CFL) phases [8]. Recent works indicate that the 2SC phase may be accessible to the two future accelerators mentioned upstream [16]. As with the hot quark gluon plasma, additional theoretical investigations will certainly be done in order to make predictions about this phase. These predictions should concern the quest of proofs of its formation during heavy ion collisions [17]. In the framework of the quark gluon plasma, these proofs involve the observation of the produced hadrons, leptons and radiations. Consequently, the knowledge of the behavior of the mesons and baryons in a color superconducting regime may present advantages. At finite densities, the fermion sign problem [18] constitutes a serious limitation of the Quantum ChromoDynamics (QCD), at least at the present state of the art. This suggests the use of effective models to study these composite particles in these conditions.

Among the various available tools, the Nambu and Jona-Lasinio (NJL) model [19,20] has proved for long its reliability to describe the quark physics at finite temperature and densities [21-29]. However, since this description considers frozen gluons, it is also known for its lack of confinement. The inclusion of a Polyakov loop [30] in this modeling has permitted to (partially) compensate this feature. This adaptation of the model constitutes the PNJL description [31-35]. The literature has reported the various advantages of this approach, like the suppression of colored states contribution at low temperatures [36]. Furthermore, an interesting feature of the PNJL model is its ability to reproduce reliability the LQCD results at vanishing densities [37]. In other words, the PNJL description may extrapolate LQCD data at finite densities. For these reasons, the PNJL model appears as an important tool to describe the color superconducting regime [38]. Moreover, in order to improve the reliability of this extrapolation, it has been proposed some improvements recently. One of them consists in including a $N_f$ correction and a $\mu$-dependence in the Polyakov loop Potential [39]. This gives the $\mu$PNJL model [40], which appears particularly relevant to describe moderate or high densities.

On the one hand, the NJL model and its various adaptations have been used to study quarks in a color superconducting regime. There are two finalities of such works. The first is to model the interior of compact stars. In these conditions, the color and charge neutrality of the quark matter must be considered. These constraints are frequently complemented with the inclusion of the $\beta$ equilibrium [7]. So, in these descriptions, the chemical potentials of the quarks and electrons are constrained by relations. The second is to explore low temperatures and high densities zones of the phase diagram. In this case, the chemical potentials of the quarks may be considered as independent parameters. An application of such investigation may concern the future accelerator programs. The Ref. [41] constitutes a review of the works performed with the NJL model in a color superconducting regime, and presents the used formalism. Furthermore, Ref. [42] underlines that the meson condensation [43,44] is another phenomenon to be taken into account in the description of the phase diagram. In parallel, Ref. [45] introduces an elegant method to handle quark/antiquark propagators. Moreover, the description of the color superconductivity with the PNJL model has been firstly proposed in Ref. [46]. The inclusion of the Polyakov loop in this regime is considered again in Refs. [16,47]. In Ref. [48], the $\mu$PNJL model has been also considered to describe the color superconductivity. In this Ref., the relevancy of the analytical method of Ref. [49] has been underlined.

On the other hand, the description of the mesons with the NJL model has been described in Refs. [22,26] and also in Refs. [27,50] for the scalar and pseudoscalar mesons. Then, such study has been extended to the PNJL model [32,34]. The main idea of the modeling consists in treating a meson as a loop formed of two quarks: a quark going towards the future and the other goes towards the past, i.e. an antiquark. The description of the diquarks follows an equivalent pattern [25,51-54]: in the loop, the charge conjugation is applied to the second quark. Consequently, this trick mimics the behavior of two real quarks, i.e. both go towards the future. The possibility to describe baryons as a quark-diquark bound state [25] constitutes a strong motivation to model diquarks. All these particles have been studied in the NJL and PNJL models, according to the temperature and the baryonic density [53,54]. Their masses are usable to intervene in cross-section calculations [55] and then in a dynamical model of evolution [56]. In the (P)NJL literature, the masses of these composite particles are frequently estimated with the method of Ref. [57]. The presented algorithms have undeniable advantages: reliability, rapidity, etc. Nevertheless, they also involve approximations, whose impact on the results should be investigated.

To describe the mesons and diquarks in a color superconducting regime, their equations must be updated, and the algorithms should be redesigned. Prior investigations of this kind are rare. However, the works of Refs. [58-62] and Refs. [63,64] present an elegant approach to describe these composite particles in, respectively, the 2SC and CFL phases. These Refs. could be considered as a starting point. In these papers, the NJL model is used at null temperature and in an $SU(2)_f$ description. Therefore, it should be instructive to extend this analysis to the $\mu_q, T$ [65,66] and $\rho_B, T$ planes. More precisely, calculations at finite temperatures and chemical potentials should allow studying the effects of the color superconductivity on the cold and dense matter. It has been observed the modest influence of the 2SC phase on the quarks masses [48]. However, this conclusion cannot be generalized to composite particles, and this aspect should be examined. Furthermore, it could be relevant to see if the meson condensation [42] could interfere with the results found in the color superconducting regime, in the $\rho_B, T$ and $\mu_q, T$ planes. In addition, a crucial development of Refs. [58-62] or Refs [65,66] could be the inclusion of the strange quarks, i.e. to work with $SU(3)_f$, in order to add $\eta, \eta'$ and the kaons in the description. Furthermore, since the PNJL model seems to become a standard in such calculations [38], the established equations should be able to take into account the inclusion of the Polyakov loop.

The Ref. [65] models mesons and diquarks in a two-color PNJL model with two flavors. However, the first order chiral phase transition is absent in this description. As a consequence, important differences are expected when three colors are involved [67]. Nevertheless, a three-color PNJL modeling of the mesons and diquarks in the 2SC phase is still missing in the literature.

In order to investigate the points mentioned upstream, I propose to organize this paper as follow. Firstly, the Sec. II focuses on a description of the quarks. In a first time, I present the formalism used to describe these particles in the color superconducting regime or when the meson condensation intervenes. More precisely, this study considers scalar diquark condensates and pseudoscalar meson condensation, i.e. the dominant pairings for these phenomena. Results are then presented. They include calculations in the $\rho_u, \rho_d$ and $\mu_u, \mu_d$ planes, without the isospin symmetry. Other results gather data obtained in the $\rho_B, T$ and $\mu_q, T$ planes, with this symmetry. The neutrality conditions and the $\beta$ equilibrium are not applied, as in Refs. [42,48]. In the Sec. III, the method employed to model mesons and diquarks in the 2SC phase is detailed. This description focuses on the pseudoscalar and scalar mesons, and idem for the diquarks. The equations are firstly given in the general case, and then simplified with the application of the isospin symmetry. I leave the study of the baryons in a superconducting regime for a future publication. The Sec. IV. shows the data found for the mesons and diquarks. This includes an analysis of their masses in the $\rho_B, T$ and $\mu_q, T$ planes. In this description, the main developed topics are: comparison of the various versions of the model (using or not the method of Ref. [57]), influence of the 2SC phase on the mesons and diquarks, stability of these particles, their behavior near the first order chiral phase transition, and modifications leaded by the inclusion of the Polyakov loop.

## II. DESCRIPTION OF THE QUARKS

### A. The NJL and ($\mu$)PNJL models

#### 1. The Lagrangian in the mean field approximation

The PNJL Lagrangian density [25,41,63,68] used in this work is expressed as

$$\begin{aligned}
\mathcal{L} = & \sum_{f=u,d,s} \bar{\psi}_f \left( i\slashed{D} - m_{0f} + \gamma_0 \mu_{0f} \right) \psi_f - \mathcal{U}(T, \Phi, \bar{\Phi}) \\
& + G \sum_{a=0}^{8} \left[ (\bar{\psi} \tau_a \psi)^2 + (\bar{\psi} i\gamma_5 \tau_a \psi)^2 \right] \\
& - G_V \sum_{a=0}^{8} \left[ (\bar{\psi} \gamma_\mu \tau_a \psi)^2 + (\bar{\psi} \gamma_\mu i\gamma_5 \tau_a \psi)^2 \right] \\
& - K \left\{ \det_f \left[ \bar{\psi}(1+\gamma_5)\psi \right] + \det_f \left[ \bar{\psi}(1-\gamma_5)\psi \right] \right\} \\
& + G_{DIQ} \sum_{a=2,5,7} \sum_{a'=2,5,7} \left( \bar{\psi} i\gamma_5 \tau_a \lambda_{a'} \psi^C \right) \left( \bar{\psi}^C i\gamma_5 \tau_a \lambda_{a'} \psi \right) \\
& + G_{DIQ} \sum_{a=2,5,7} \sum_{a'=2,5,7} \left( \bar{\psi} \tau_a \lambda_{a'} \psi^C \right) \left( \bar{\psi}^C \tau_a \lambda_{a'} \psi \right)
\end{aligned} \qquad (1)$$

In this writing, $\psi = (\psi_u, \psi_d, \psi_s)^T$ refers to the quark fields. Since the Polyakov loop is included, $\slashed{D} = \gamma^\mu D_\mu$ uses the derivative $D_\mu = \partial_\mu - iA_\mu$, with $A_0 = -iA_4$, and the effective Polyakov loop potential $\mathcal{U}(T, \Phi, \bar{\Phi})$ is present [32,34]. These terms are detailed Sect. II.A.3. Also, $\mu_{0f}$ is the bare chemical potential of the flavor $f$ quarks and $m_{0f}$ is their naked mass. $G$, $G_V$, $K$ correspond to the terms that describe, respectively, the scalar quark-antiquark interaction, the vector interaction in order to work at finite densities, and the 't Hooft term. In addition, $G_{DIQ}$ is related to the scalar and pseudoscalar quark-quark interaction terms. Axial and vector interactions between two quarks are not considered. With the cutoff $\Lambda$, all the mentioned constants constitute a parameter set. In this work, two sets are considered, named P1 and EB, whose values appear in TABLE I [48,53-56]. The first is designed for calculations that respect the isospin symmetry, leading to $m_u = m_d \equiv m_q$ and $\mu_u = \mu_d \equiv \mu_q$. In contrast, the second is employed when these two relations are not applied.

**TABLE I.** Parameter sets used in the (P)NJL models. The masses $m_{0f}$ and the cutoff $\Lambda$ are in MeV, $G$, $G_V$ and $G_{DIQ}$ are in MeV$^{-2}$ and $K$ is in MeV$^{-5}$.

| Parameter set | $m_{0u}$ | $m_{0d}$ | $m_{0s}$ | cutoff $\Lambda$ | $G\Lambda^2$ | $K\Lambda^5$ | $G_V$ | $G_{DIQ}$ |
|---|---|---|---|---|---|---|---|---|
| P1 | 4.75 | 4.75 | 147.0 | 708.0 | 1.922 | 10.00 | 0.310 $G$ | 0.705 $G$ |
| EB | 4.00 | 6.00 | 120.0 | 708.0 | 1.922 | 10.00 | 0.295 $G$ | 0.705 $G$ |

In Eq. (1), $\tau_j$ and $\lambda_k$ are Gell-Mann matrices: divided by a factor 2, they are the 8 generators of, respectively, $SU(3)_{flavor}$ and $SU(3)_{color}$. Also, $\lambda_0 = \sqrt{2/3}\, 1_3$ is added, where $1_3$ is the $3 \times 3$ identity matrix. $\mathcal{C} = i\gamma^2\gamma^0$ is the charge conjugation operator and explains why the fifth line translate a *scalar* interaction, and the sixth a *pseudoscalar* one [25]. These quark-quark interactions are antisymmetric in flavor and color, which justifies the use of the antisymmetric matrices $\tau_{2,5,7}$ and $\lambda_{2,5,7}$.

Then, the mean field approximation is applied to the Eq. (1). At this occasion, the shorthand notation $\sigma_f = \langle\langle \bar{\psi}_f \psi_f \rangle\rangle$ is introduced to designate the chiral condensate and $\rho_f = \langle\langle \psi_f^+ \psi_f \rangle\rangle$ corresponds to the density of the flavor $f$ quarks. With the Hartree approximation, if the pseudoscalar diquark and scalar meson condensates are neglected, the Lagrangian density becomes [7,41,61,69,70]

$$\begin{aligned}\mathcal{L}_{MF} = &\sum_{f=u,d,s} \bar{\psi}_f \left[ i\slashed{D} - \left(m_{0f} - 4G\sigma_f + 2K\sigma_j\sigma_k\right) + \gamma_0 \left(\mu_{0f} - 4G_V \rho_f\right)\right]\psi_f \\ &- \mathcal{U}(T,\Phi,\bar{\Phi}) + \bar{\psi}\left[-i\gamma_5 \tau_2 \Delta_\pi - i\gamma_5 \tau_5 \Delta_K - i\gamma_5 \tau_7 \Delta_{K^0}\right]\psi \\ &+ \bar{\psi}^{\mathcal{C}}\left[-\gamma_5 \tau_2 \lambda_2 \frac{\Delta^*_{ud}}{2} - \gamma_5 \tau_5 \lambda_5 \frac{\Delta^*_{us}}{2} - \gamma_5 \tau_7 \lambda_7 \frac{\Delta^*_{ds}}{2}\right]\psi \\ &+ \bar{\psi}\left[\gamma_5 \tau_2 \lambda_2 \frac{\Delta_{ud}}{2} + \gamma_5 \tau_5 \lambda_5 \frac{\Delta_{us}}{2} + \gamma_5 \tau_7 \lambda_7 \frac{\Delta_{ds}}{2}\right]\psi^{\mathcal{C}} \\ &- 2G \sum_{f=u,d,s} \sigma_f^2 - \frac{1}{4G}\left(\Delta_\pi^2 + \Delta_K^2 + \Delta_{K^0}^2\right) \\ &+ 4K\sigma_u \sigma_d \sigma_s + 2G_V \sum_{f=u,d,s} \rho_f^2 - \frac{1}{4G_{DIQ}}\left(|\Delta_{ud}|^2 + |\Delta_{us}|^2 + |\Delta_{ds}|^2\right)\end{aligned} \quad (2)$$

The first line of this relation allows identifying the gap equation for the quark masses

$$m_f = m_{0f} - 4G\sigma_f + 2K\sigma_j\sigma_k \Big|_{\substack{f=u,d,s \\ f\neq j \text{ and } f\neq k}}. \quad (3)$$

In other words, $m_f$ is the dressed mass of a flavor $f$ quark. In the same way,

$$\mu_f = \mu_{0f} - 4G_V \rho_f, \quad (4)$$

where $\mu_f$ is considered as an effective chemical potential. In the other lines of Eq. (2), the following relations are employed to introduce the energy gaps of the pseudoscalar meson condensates [42,69]

$$\Delta_\pi = -2G\langle\langle\bar{\psi} i\gamma_5 \tau_2 \psi\rangle\rangle, \quad \Delta_K = -2G\langle\langle\bar{\psi} i\gamma_5 \tau_5 \psi\rangle\rangle, \quad \Delta_{K^0} = -2G\langle\langle\bar{\psi} i\gamma_5 \tau_7 \psi\rangle\rangle \quad (5)$$

and of the scalar diquark condensates

$$\Delta_{ud} = -2G_{DIQ} \left\langle \left\langle \overline{\psi}^C \gamma_5 \tau_2 \lambda_2 \psi \right\rangle \right\rangle, \quad \Delta_{us} = -2G_{DIQ} \left\langle \left\langle \overline{\psi}^C \gamma_5 \tau_5 \lambda_5 \psi \right\rangle \right\rangle, \quad \Delta_{ds} = -2G_{DIQ} \left\langle \left\langle \overline{\psi}^C \gamma_5 \tau_7 \lambda_7 \psi \right\rangle \right\rangle, \quad (6)$$

with also $\Delta^*_{ff'} = \left(\Delta_{ff'}\right)^\dagger$ [59]. In the normal quark phase (NQ), these six gaps are equal to zero. In the 2SC phase, only $\Delta_{ud} \neq 0$. In the CFL phase, $\Delta_{ud}$, $\Delta_{us}$ and $\Delta_{ds}$ are non-null.

## 2. The grand potential

Using the standard techniques, the Hamiltonian is then obtained from $\mathcal{L}_{MF}$. It allows finding the expression of the grand potential [48,68]

$$\begin{aligned}
\Omega = &\Omega_M + \mathcal{U}(T, \Phi, \overline{\Phi}) \\
&+ 2G \sum_{f=u,d,s} \left\langle \left\langle \overline{\psi}_f \psi_f \right\rangle \right\rangle^2 + G \left( \left\langle \left\langle \overline{\psi}_u \gamma_5 \psi_d \right\rangle \right\rangle^2 + \left\langle \left\langle \overline{\psi}_s \gamma_5 \psi_u \right\rangle \right\rangle^2 + \left\langle \left\langle \overline{\psi}_s \gamma_5 \psi_d \right\rangle \right\rangle^2 \right) \\
&- 4K \left\langle \left\langle \overline{\psi}_u \psi_u \right\rangle \right\rangle \left\langle \left\langle \overline{\psi}_d \psi_d \right\rangle \right\rangle \left\langle \left\langle \overline{\psi}_s \psi_s \right\rangle \right\rangle - 2G_V \sum_{f=u,d,s} \left\langle \left\langle \psi_f^+ \psi_f \right\rangle \right\rangle^2 \\
&+ G_{DIQ} \left( \left\langle \left\langle \psi_u \psi_d \right\rangle \right\rangle^2 + \left\langle \left\langle \psi_u \psi_s \right\rangle \right\rangle^2 + \left\langle \left\langle \psi_u \psi_s \right\rangle \right\rangle^2 \right)
\end{aligned}, \quad (7)$$

with the shorthand notations $\left\langle \left\langle \overline{\psi}_u \gamma_5 \psi_d \right\rangle \right\rangle \equiv \left\langle \left\langle \overline{\psi} i \gamma_5 \tau_2 \psi \right\rangle \right\rangle$ [71], $\left\langle \left\langle \psi_u \psi_d \right\rangle \right\rangle \equiv \left\langle \left\langle \overline{\psi}_u^C \gamma_5 \psi_d \right\rangle \right\rangle$ [48] and so on for the other flavors. The thermodynamical potential $\Omega_M$ of the $u, d, s$ quarks/antiquarks is expressed as [41,68]

$$\Omega_M = -\frac{T}{2} \int \frac{d^3 \vec{p}}{(2\pi)^3} \sum_n \text{Tr} \left\{ \ln \left[ \beta \tilde{S}^{-1}(i\omega_n, \vec{p}) \right] \right\}. \quad (8)$$

In Eq. (8), $T = 1/\beta$ is the temperature. The matrix $\tilde{S}^{-1}$ is the inverse propagator of the quarks/antiquarks. In the Nambu-Gorkov basis $\Psi = (\psi, \psi^C)^T$, this matrix is written as [7,41]

$$\tilde{S}^{-1} = \begin{bmatrix} \left(S_0^+\right)^{-1} & \Delta^- \\ \Delta^+ & \left(S_0^-\right)^{-1} \end{bmatrix}. \quad (9)$$

When the pseudoscalar meson condensation is taken into account [42,65,66,71], the matrices

$$\left(S_0^\pm\right)^{-1} = \begin{bmatrix} \slashed{p} \pm \gamma_0 \tilde{\mu}_u - m_u & \mp \gamma_5 \Delta_\pi & \mp \gamma_5 \Delta_K \\ \pm \gamma_5 \Delta_\pi & \slashed{p} \pm \gamma_0 \tilde{\mu}_d - m_d & \mp \gamma_5 \Delta_{K^0} \\ \pm \gamma_5 \Delta_K & \pm \gamma_5 \Delta_{K^0} & \slashed{p} \pm \gamma_0 \tilde{\mu}_s - m_s \end{bmatrix}, \quad (10)$$

reveal the corresponding couplings in the non-diagonal terms. In $\slashed{p} \pm \gamma_0 \tilde{\mu}_f - m_f$, the $\tilde{\mu}_f$ are $3 \times 3$ matrices that gather $N_c = 3$ chemical potentials:

$$\tilde{\mu}_f = \text{diag}\left(\mu_f^r, \mu_f^g, \mu_f^b\right) = \mu_f 1_3 - iA_4. \quad (11)$$

Furthermore, the matrices $\Delta^\mp$ describe the quark-quark pairings that intervene in the color superconducting regime. In a description involving the scalar coupling (the dominant one) [14,41,68],

$$\Delta^- = \Delta_{ff',cc'} \gamma_5 \otimes \tau_{ff'} \otimes \lambda_{cc'},$$

which gives

$$\Delta^- = \begin{bmatrix} 0 & -i\gamma_5 \Delta_{ud} \lambda_2 & -i\gamma_5 \Delta_{us} \lambda_5 \\ i\gamma_5 \Delta_{ud} \lambda_2 & 0 & -i\gamma_5 \Delta_{ds} \lambda_7 \\ i\gamma_5 \Delta_{us} \lambda_5 & i\gamma_5 \Delta_{ds} \lambda_7 & 0 \end{bmatrix} \text{ and } \Delta^+ = \gamma_0 \left(\Delta^-\right)^\dagger \gamma_0. \quad (12)$$

In practice, the matrix $\tilde{S}$ can be put on the form [7]

$$\tilde{S} = \begin{bmatrix} S^+ & \Xi^- \\ \Xi^+ & S^- \end{bmatrix}, \qquad (13)$$

where the quark/antiquark propagators satisfy the property [15]

$$S^{\pm} = \left[ \left( S_0^{\pm} \right)^{-1} - \Delta^{\mp} S_0^{\mp} \Delta^{\pm} \right]^{-1} \qquad (14)$$

and the abnormal propagators are found with

$$\Xi^{\pm} = -S^{\mp} \Delta^{\pm} S_0^{\pm}. \qquad (15)$$

Appendix A gathers the various expressions of these propagators according to the treated phase. In addition, Appendix B proposes an alternative writing, which is obtained with the method of Ref. [45], in the framework of the isospin symmetry.

### 3. Detail on the Polyakov loop

As observed in the previous Sect., the inclusion of the Polyakov loop leads to consider the temporal component $A_4$ of the Euclidean gauge field $A$ [32]. In the Polyakov gauge, this $3 \times 3$ matrix can be written as $A_4 = \text{diag}\left( A_{4(11)}, A_{4(22)}, A_{4(33)} \right)$ [31]. Consequently, the chemical potentials of Eq. (11) are writable with the shorthand notation

$$\mu_f^r = \mu_f - iA_{4(11)}, \quad \mu_f^g = \mu_f - iA_{4(22)}, \quad \mu_f^b = \mu_f - iA_{4(33)} \qquad (16)$$

Obviously, $\mu_f^r = \mu_f^g = \mu_f^b \equiv \mu_f$ in the NJL description, i.e. without the Polyakov loop. Furthermore, $A_4$ can be defined as $\beta A_4 = \phi_3 \lambda_3 + \phi_8 \lambda_8$ [33], where $\phi_3, \phi_8$ are real numbers. $A_4$ is usable to define the Polyakov line

$$L(\vec{x}) = \mathcal{P} \exp\left( i \int_0^{\beta} A_4(\vec{x}, \tau) \, d\tau \right) \qquad (17)$$

and its conjugate $L^{\dagger}$, where $\mathcal{P}$ is a path ordering operator. $L = \exp(i\beta A_4)$ and $L^{\dagger} = \exp(-i\beta A_4)$ in the mean field approximation. In addition, one defines $\Phi$ and $\bar{\Phi}$ as, respectively, the average of the Polyakov field and its complex conjugate

$$\Phi = \frac{1}{N_c} \text{Tr}_c(L), \quad \bar{\Phi} = \frac{1}{N_c} \text{Tr}_c\left( L^{\dagger} \right) \qquad (18)$$

in which $\text{Tr}_c$ is a trace over the $N_c$ colors. In pure gauge calculations, $\Phi$ is an order parameter of the $\mathbb{Z}_3$ symmetry. It indicates the confined regime when $\Phi = 0$ and the deconfined regime when $\Phi \to 1$. When dynamical quarks are included, the $\mathbb{Z}_3$ symmetry breaking is no longer exact. In these conditions, the words "confined" or "confined" are still employed in PNJL papers, but often with quotation marks. Indeed, the inclusion of the Polyakov loop only mimics a mechanism of confinement in the PNJL description [32,48].

Another consequence of the inclusion of the Polyakov loop is the apparition of the Polyakov loop potential $\mathcal{U}(T, \Phi, \bar{\Phi})$ in the Lagrangian density, Eq. (1). In the literature, the two most encountered expressions of this potential are composed of powers of $\Phi, \bar{\Phi}$ [72,73], or of a logarithm term [33,46]. In this paper, the second is employed because it allows keeping $\Phi < 1$ in all the cases. Its expression is

$$\frac{\mathcal{U}(T, \Phi, \bar{\Phi})}{T^4} = -\frac{a(T)}{2} \Phi \bar{\Phi} + b(T) \ln\left[ 1 - 6\Phi\bar{\Phi} + 4\left( \Phi^3 + \bar{\Phi}^3 \right) - 3\left( \Phi\bar{\Phi} \right)^2 \right]. \qquad (19)$$

$a(T) = a_0 + a_1 (T_0/T) + a_2 (T_0/T)^2$ and $b(T) = b_3 (T_0/T)^3$, with $a_0 = 3.51$, $a_1 = -2.47$, $a_2 = 15.2$ and $b_3 = -1.75$. In pure gauge calculations, $T_0$ corresponds to the critical temperature of the deconfinement transition. In the usual PNJL description, this temperature is constant and $T_0 = 270$ MeV [31]. In contrast, when a quark back-reaction to the gluonic sector is included [39,40,48] ($\mu$ PNJL model), $T_0$ is rewritten as

$$T_0 = T_\tau \exp\left[\frac{-1}{\alpha_0 b(N_f, \mu_f)}\right] \text{ where } b(N_f, \mu_f) = \frac{11N_c - 2N_f}{6\pi} - \frac{16}{\pi}\sum_f \left(\frac{\mu_f}{T_\tau}\right)^2 \quad (20)$$

in order to include $N_f$ corrections and the $\mu$-dependence in $\mathcal{U}(T, \Phi, \bar{\Phi})$. In this relation, $T_\tau = 1770$ MeV, $\alpha_0 = 0.304$ and $N_f$ is the number of (approximately) massless flavors.

*4. Resolution of the equations*

The estimation of the dressed quark masses is an important stage of the modeling. In all the cases, it requires solving $m_f = m_{0f} - 4G\sigma_f + 2K\sigma_j\sigma_k$ Eq. (3) for each flavor: this forms a system of equations. Nevertheless, the analytical expression of the chiral condensates $\sigma_f = \langle\langle\bar{\psi}_f\psi_f\rangle\rangle$ depend on the treated phase [48] and is found with one of the three relations [15,27,41]

$$\frac{\partial\Omega}{\partial\langle\langle\bar{\psi}_f\psi_f\rangle\rangle} = 0, \quad \langle\langle\bar{\psi}_f\psi_f\rangle\rangle = \frac{\partial\Omega_M}{\partial m_f} \quad \text{or} \quad \langle\langle\bar{\psi}_f\psi_f\rangle\rangle = \frac{1}{\beta}\sum_n \int\frac{d^3\vec{p}}{(2\pi)^3}\text{Tr}(S_f^+). \quad (21)$$

When the calculations are performed according to the densities, $\rho_f = \langle\langle\psi_f^+\psi_f\rangle\rangle$ must be included in the system of equations. For each flavor, $\rho_f$ corresponds to the wanted density. Also, the expression of $\langle\langle\psi_f^+\psi_f\rangle\rangle$ is obtained with [26,41,74]

$$\frac{\partial\Omega}{\partial\langle\langle\psi_f^+\psi_f\rangle\rangle} = 0, \quad \langle\langle\psi_f^+\psi_f\rangle\rangle = -\frac{\partial\Omega_M}{\partial\mu_f} \quad \text{or} \quad \langle\langle\psi_f^+\psi_f\rangle\rangle = \frac{1}{\beta}\sum_n \int\frac{d^3\vec{p}}{(2\pi)^3}\text{Tr}(\gamma_0 S_f^+). \quad (22)$$

In the normal quark (NQ) phase described by an $SU(3)_f$ NJL model, the masses and the chemical potentials of the quarks are the six unknowns of this six-equation system. The temperature $T$ and the densities $\rho_f$ are parameters. Nevertheless, the isospin symmetry allows obtaining a system of four equations with four unknowns, i.e. $m_q, m_s, \mu_q$ and $\mu_s$.

The inclusion of the Polyakov loop leads to two extra relations in the system of equations [33,46]

$$\frac{\partial\text{Re}(\Omega)}{\partial\phi_3} = 0 \quad \text{and} \quad \frac{\partial\text{Re}(\Omega)}{\partial\phi_8} = 0. \quad (23)$$

The $\phi_3$, $\phi_8$ are two extra unknowns. The simplification that consists in neglecting $\text{Im}(\Omega)$ is discussed in Refs. [33,75,76]. Moreover, in the PNJL literature,

$$\partial\Omega/\partial\Phi = 0 \quad \text{and} \quad \partial\Omega/\partial\bar{\Phi} = 0 \quad (24)$$

are also observable. They treat $\Phi$ and $\bar{\Phi}$ as real and independent variables. This approximation is only employable in the NQ phase [46,48]. In other words, the "standard" ($\mu$)PNJL model [34,40,54] uses Eq. (24), whereas the ($\mu$)PNJL versions that describe the color superconductivity [33,46,48] or the meson condensation consider Eq. (23). These versions will be labeled as $\Delta(\mu)$PNJL [48]. More generally, the $\Delta$ placed in front of the model's name recalls that the used equations can include the color superconductivity (or the meson condensation in Sec. II.B.1). In this paper, this remark will concern the description of the quarks, mesons and diquarks.

To model the color superconducting phases, one gap equation must be added for each energy gap $\Delta_{ff'}$. This equation is established with one of the relations [15,41,74]

$$\frac{\partial \Omega}{\partial |\Delta_{ff'}|} = 0 \quad \text{or} \quad \left|\Delta_{ff'}^{cc'}\right| = 2G_{DIQ} \frac{1}{\beta} \sum_n \int \frac{d^3\vec{p}}{(2\pi)^3} \left|\text{Tr}\left(\Xi_{ff'}^{cc'-} \gamma_5\right)\right|. \tag{25}$$

Concerning the description of the pseudoscalar meson condensation, each gap $\Delta_\pi, \Delta_K, \Delta_{K^0}$ is associated with one equation. The relation for the pion condensate is found with

$$\frac{\partial \Omega}{\partial \Delta_\pi} = 0, \quad \Delta_\pi = -2G \frac{\partial \Omega_M}{\partial \Delta_\pi} \quad \text{or} \quad \Delta_\pi = -2G \frac{1}{\beta} \sum_n \int \frac{d^3\vec{p}}{(2\pi)^3} \text{Tr}\left(S_f^+ \gamma_5\right), \tag{26}$$

and so on for the other gaps. When all of them are null except for $\Delta_\pi$, one gets

$$1 = 8GN_c \int \frac{d^3\vec{p}}{(2\pi)^3} \frac{1}{\sqrt{E_q^2 + \Delta_\pi^2}} \left[1 - f_{FD}\left(\sqrt{E_q^2 + \Delta_\pi^2} + \mu_q\right) - f_{FD}\left(\sqrt{E_q^2 + \Delta_\pi^2} - \mu_q\right)\right] \tag{27}$$

in the NJL description applying the isospin symmetry. $f_{FD}(x) = \left(e^{\beta x} + 1\right)^{-1}$ is the Fermi-Dirac distribution, and $E_q^2 = \vec{p}^2 + m_q^2$.

The relations that form the system of equations are coupled [48]: Eq. (27) needs $m_q$ obtained with Eq. (3), and conversely Eq. (3) could require $\Delta_\pi$ found with Eq. (27)… Consequently, these equations are solved simultaneously. Once this is done, the found values (masses, chemical potentials, Polyakov fields and energy gaps) are then fully usable to model the mesons and diquarks, via the equations established in Sec. III.

**B. Results**

*1. Pseudoscalar meson condensation and color superconductivity at low temperatures*

The meson condensation and the color superconductivity have been studied together in Ref. [42], via phase diagrams according to the chemical potentials. An objective of this Sect. is to find again such results with the EB parameter set, i.e. without the isospin symmetry. Another goal is to perform this analysis according to the densities, which is rather new for the meson condensation.

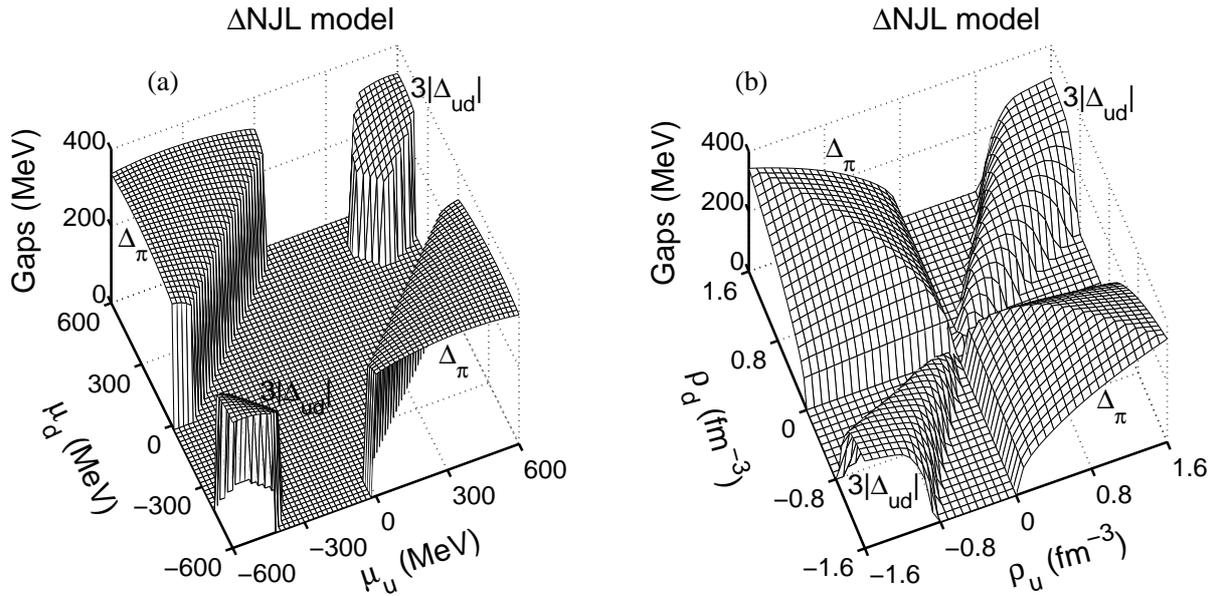

**FIG 1.** Evolution of the gaps $\Delta_{ud}$ and $\Delta_\pi$ in the (a) $\mu_u, \mu_d$ and (b) $\rho_u, \rho_d$ planes.

The strangeness is fixed to zero in the graphs exhibited in this paper. Therefore, the calculations focus on a description of the 2SC phase ($\Delta_{ud} \neq 0$) and the pion condensation ($\Delta_\pi \neq 0$). The results of this Sec. are found with the $\Delta$ NJL model, at $T = 5$ MeV. This temperature is low enough to consider that the $\Delta$ PNJL and $\Delta\mu$ PNJL descriptions would have given very similar graphs.

Firstly, I have verified that the modeling described in the previous Sects. allows finding again the data of Ref. [42] with its parameter set. Then, I have established FIG 1 with the EB parameter set. Qualitatively, FIG 1(a) is very close to the first graph of this Ref. However, the main difference is the width of the "corridor" observable between the two $\Delta_\pi \neq 0$ zones. This width is greater with the EB parameter set, and may be explained by the inclusion of the 't Hooft term.

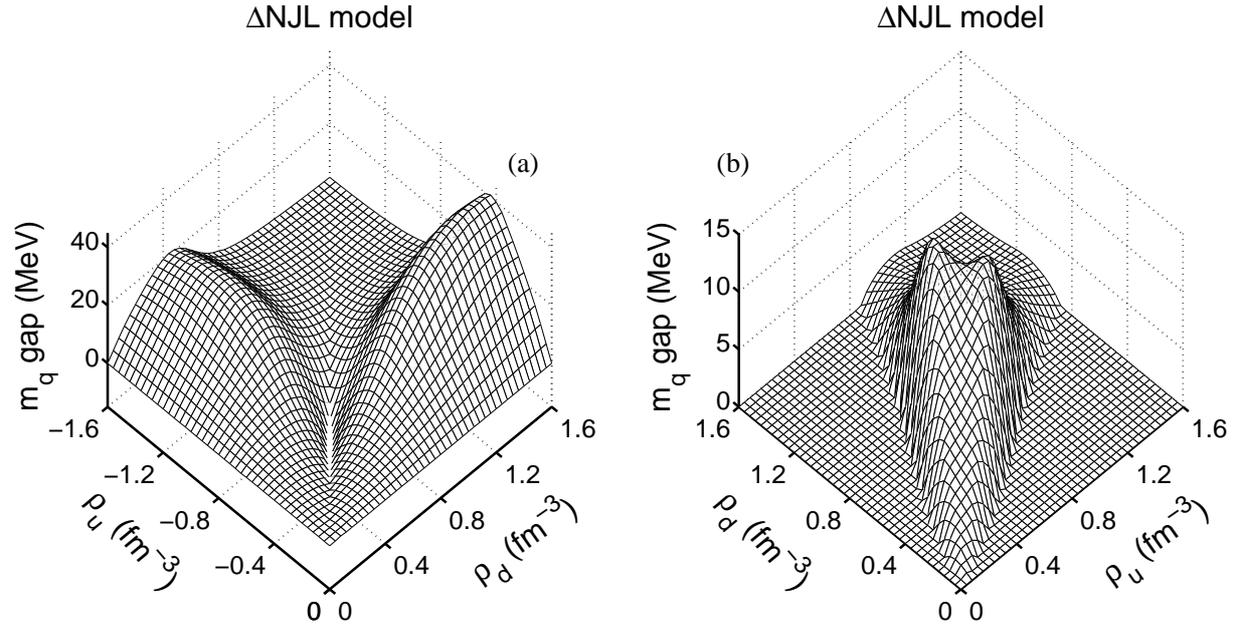

**FIG 2.** Increase in the mass $m_q$ of the light quarks caused by the (a) pion condensation and the (b) 2SC phase.

The intensity of the pairings is another aspect revealed by FIG 1. The gap $\Delta_\pi$ can be largely stronger than $|\Delta_{ud}|$. Consequently, the influence of the pion condensation on some observables is more important than the one of the 2SC phase. This affirmation is confirmed by FIG 2, which studies the effects of the mentioned phases on the quark masses. Moreover, FIG 1 shows that the pion condensation and the 2SC regime cannot interfere. Indeed, there is no overlap between them in the $\mu_u, \mu_d$ and $\rho_u, \rho_d$ planes, and they do not have common frontier. Nevertheless, as recalled by Ref. [42], this observation cannot be generalized, especially at finite strangeness: the kaon condensation and 2SC phases are separated by a first order phase transition. Furthermore, the pion condensation only intervenes when $\mu_u$ and $\mu_d$ are very different, and idem for the densities. So, there is no possible influence of the pion condensation when the isospin symmetry is respected. Since the other calculations described in this paper considers this symmetry, one focuses on a description of the mesons and diquarks in the NQ and 2SC phases.

## 2. Overview on some 2SC results

The quarks have been studied in the $\rho_B, T$ and $\mu_q, T$ planes at several occasions with the NJL model or its various versions (cf. Refs in the introduction), where $\rho_B = \frac{2}{3}\rho_q$. In this Sect., I propose to recall results that appear relevant in the framework of the modeling performed in this paper. These data and the following ones have been obtained with the P1 parameter set.

The surface that represents the light quark mass evolves smoothly in the $\rho_B, T$ plane [77]. The inclusion of the 2SC phase in the description does not alter this feature, FIG 3(a). Indeed, this phase only leads to a slight deformation of the surface. Also, the gap $\Delta_{ud}$ does not have discontinuity in this plane, FIG 3(b). Only the second order phase transition according to the temperature is visible, when $\Delta_{ud}$ falls to zero continuously. The first order chiral phase transition, which occurs at low and moderate $T$ [70], is totally invisible in both graphs. The unstable regime associated with this transition satisfies the property $\partial^2 \Omega / \partial \sigma_q^2 < 0$. Its limits are superimposed to the plotted surfaces. In order to not weigh down the description, the metastable regimes are not included. In the framework of the *problem of matter stability* [78] and with the quark droplet picture [79], the physical sense of these unstable and metastable regimes (forming the *mixed phase*) has been discussed in Ref. [48]. Thanks to the quark droplet description, this Ref. has also explained that the presence of a color superconducting phase at very low densities reveals the lack of confinement in the $\Delta$(P)NJL modeling.

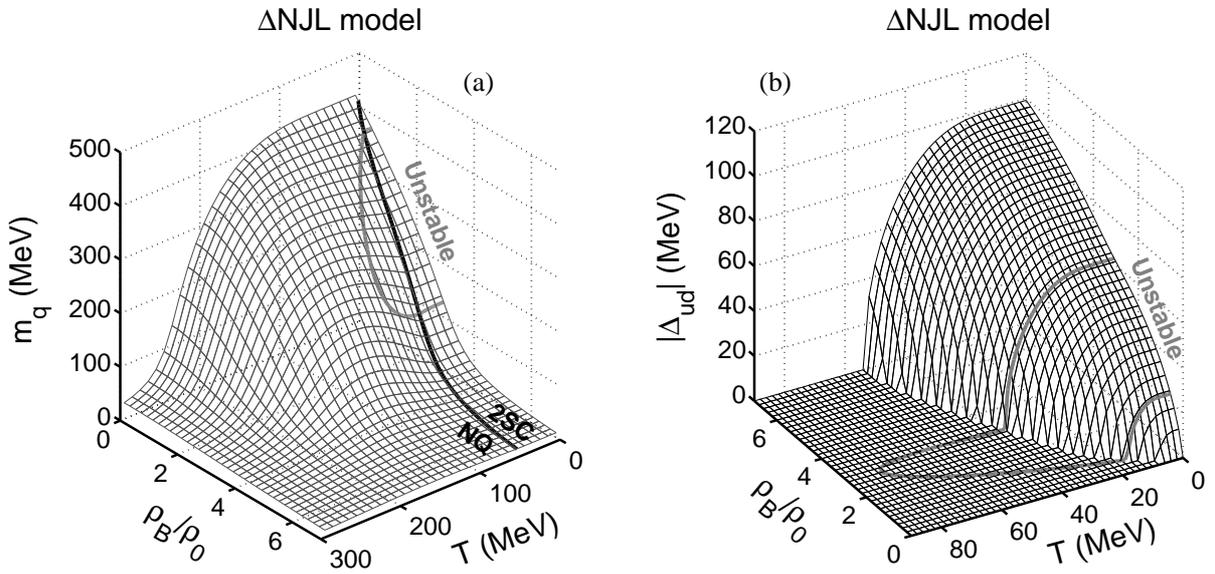

**FIG 3.** (a) Mass $m_q$ and (b) gap $\Delta_{ud}$ in the $T, \rho_B$ planes, with $\rho_0 \approx 0.16$ fm$^{-3}$.

Moreover, the data that permitted to draw FIG 3(a) are usable to represent $m_q$ in the $T, \mu_q$ plane. With the method detailed in Ref [48], the obtained surface is formed of isotemperature and isodensity lines. Furthermore, all the states (stable, unstable and metastable) are visible. So, the discontinuity due to the chiral restoration is replaced by the structure visible in the right hand side of FIG 4(a). A zoom in this structure, FIG 4(b), reveals a first order phase transition like those found in the handbooks. At low $T$, the chiral restoration and the NQ/2SC phase transition occurs at the same critical chemical potential $\mu_c \approx 395$ MeV, at least with the used parameter set. See Ref. [80] for a counterexample. In fact, the 2SC phase tends to lower $\mu_c$ by about 5 MeV. Indeed, for higher temperatures, as in FIG 4(b), this coincidence is lost and the chiral restoration intervenes at $\mu_c \approx 400$ MeV.

When the Polyakov loop is included, all the previous observations stay valid, at least qualitatively. Indeed, in the $\Delta$PNJL model, the results are not modified but shifted towards higher $T$ [53,54], FIG 5(a). This statement includes the unstable zone and the NQ-2SC frontier. This constitutes a typical feature of this description. In contrast, the $\Delta\mu$PNJL and $\Delta$NJL graphs are rather close, FIG 5(b). The main differences between them are found at low densities and at temperatures around 150 MeV. Moreover, in FIG 3(a) and FIG 5, the critical endpoint (CEP) is located at the maximum temperature reached by the unstable zone. $T_{CEP} \approx 79$ MeV in the $\Delta$NJL approach, vs. $T_{CEP} \approx 158$ MeV with

$\Delta$ PNJL and $T_{CEP} \approx 77$ MeV with $\Delta\mu$ PNJL. However, in these three versions of the model, the CEP is found at $\rho_B \approx 2.8\rho_0$. In addition, when $T$ exceeds the $T_{CEP}$, the chiral restoration is then performed via a crossover, whatever the version.

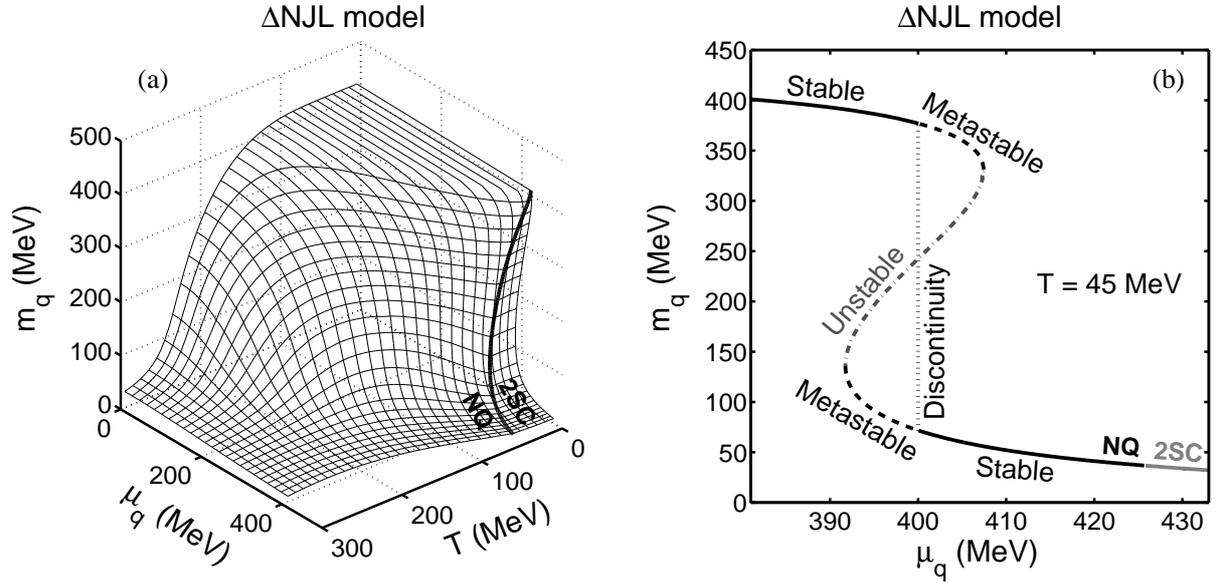

**FIG 4.** (a) Mass $m_q$ and in the $T, \mu_q$ planes and (b) detail on the curve at $T = 45$ MeV.

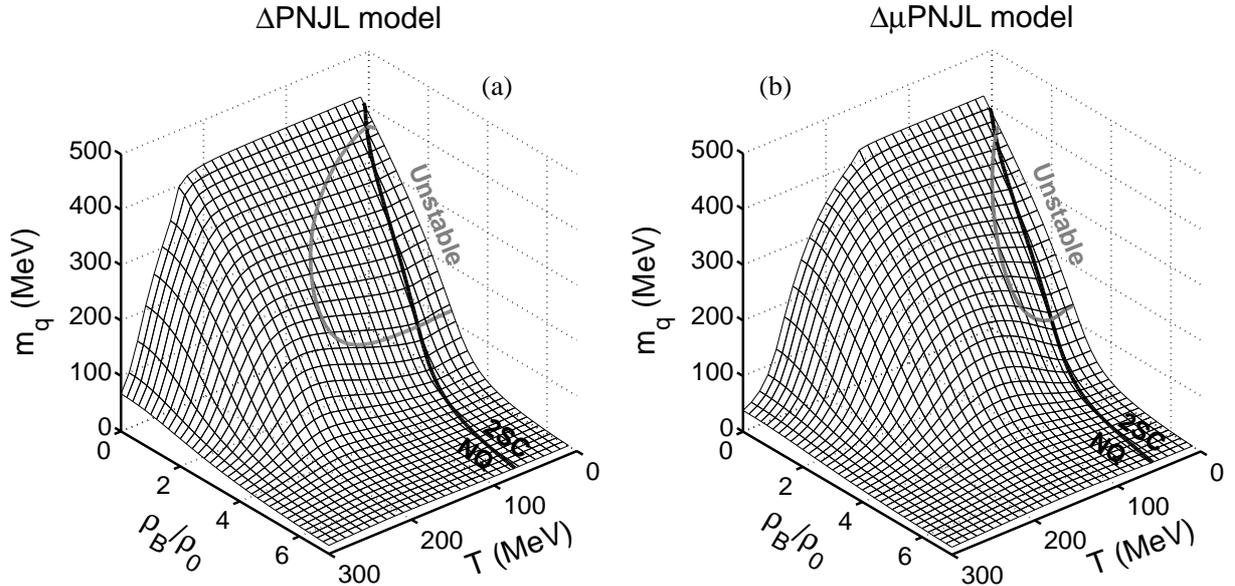

**FIG 5.** Mass $m_q$ in the $T, \rho_B$ planes, in the (a) $\Delta$ PNJL and (b) $\Delta\mu$ PNJL models.

### III. MODELING THE MESONS AND DIQUARKS

#### A. Global description of the approach

In Refs. [58-64], a method is presented to describe the mesons and diquarks in a superconducting regime. The main idea consists in considering the fluctuations of the fields around their mean values. In order to distinguish a field and its fluctuation, the latter is underlined, as in Eq. (28) [61].

Firstly, the following term is considered

$$\frac{i}{4} Tr \left( \tilde{S} \begin{bmatrix} \underline{\Sigma} & \underline{\Delta}^- \\ \underline{\Delta}^+ & \underline{\Sigma}^T \end{bmatrix} \tilde{S} \begin{bmatrix} \underline{\Sigma} & \underline{\Delta}^- \\ \underline{\Delta}^+ & \underline{\Sigma}^T \end{bmatrix} \right), \tag{28}$$

in which $Tr$ gathers traces in the Nambu-Gorkov, spinor, flavor and color spaces, and an integral over space-time coordinates [58-60,62]. In addition, $\tilde{S} = \begin{bmatrix} S^+ & \Xi^- \\ \Xi^+ & S^- \end{bmatrix}$, Eq. (13), with $S^{\pm} = \text{diag}\left(S_u^{\pm}, S_d^{\pm}, S_s^{\pm}\right)$ and $S_f^{\pm} = \text{diag}\left(S_f^{r\pm}, S_f^{g\pm}, S_f^{b\pm}\right)$ when the meson condensation is not considered. In the 2SC phase,

$$\Xi^{\pm} = \begin{bmatrix} 0 & \Xi_{ud}^{\pm} & 0 \\ \Xi_{du}^{\pm} & 0 & 0 \\ 0 & 0 & 0 \end{bmatrix}, \text{ with } \Xi_{ud}^{\pm} = \begin{bmatrix} 0 & \Xi_{ud}^{rg\pm} & 0 \\ \Xi_{ud}^{gr\pm} & 0 & 0 \\ 0 & 0 & 0 \end{bmatrix} \text{ and so on for } \Xi_{du}^{\pm}. \tag{29}$$

$\underline{\Delta}^{\pm}$ gathers the field fluctuations associated with the diquark condensates. With the scalar diquarks, one gets [63,64]

$$\underline{\Delta}^- = -i\gamma_5 \sum_{A=2,5,7} \lambda_A \begin{bmatrix} 0 & \underline{\Delta}_{ud}^A & \underline{\Delta}_{us}^A \\ -\underline{\Delta}_{ud}^A & 0 & \underline{\Delta}_{ds}^A \\ -\underline{\Delta}_{us}^A & -\underline{\Delta}_{ds}^A & 0 \end{bmatrix} \text{ and } \underline{\Delta}^+ = i\gamma_5 \sum_{A=2,5,7} \lambda_A \begin{bmatrix} 0 & \underline{\Delta}_{ud}^{A*} & \underline{\Delta}_{us}^{A*} \\ -\underline{\Delta}_{ud}^{A*} & 0 & \underline{\Delta}_{ds}^{A*} \\ -\underline{\Delta}_{us}^{A*} & -\underline{\Delta}_{ds}^{A*} & 0 \end{bmatrix}, \tag{30}$$

with $\underline{\Delta}_{ff'}^2 \equiv \underline{\Delta}_{ff'}^{rg}$, $\underline{\Delta}_{ff'}^5 \equiv \underline{\Delta}_{ff'}^{rb}$ and $\underline{\Delta}_{ff'}^7 \equiv \underline{\Delta}_{ff'}^{gb}$. The writing is similar with the pseusoscalar diquarks if $i\gamma_5$ is dropped. Moreover, the terms associated with the mesons are found in the $\underline{\Sigma}$ and $\underline{\Sigma}^T$ matrices, which are expressed as [63,64]

$$\underline{\Sigma} = -\sum_{a=0}^{8} \underline{\pi}_a \lambda_a i\gamma_5 + \underline{\sigma}_a \lambda_a 1_4 \text{ and } \underline{\Sigma}^T = -\sum_{a=0}^{8} \underline{\pi}_a \lambda_a^T i\gamma_5 + \underline{\sigma}_a \lambda_a^T 1_4, \tag{31}$$

where $^T$ is the transposition operation. The terms involving the $4 \times 4$ identity matrix $1_4$ describe scalar mesons. Those with $i\gamma_5$ concern pseudoscalar ones and are rewritable as [59]

$$\sum_{a=0}^{8} \underline{\pi}_a \lambda_a = \underline{\Delta}_0 \lambda_0 + \underline{\Delta}_{\pi^{\mp}} \left(\frac{\lambda_1 \pm i\lambda_2}{\sqrt{2}}\right) + \underline{\Delta}_{\pi^0} \lambda_3 + \underline{\Delta}_{K^{\mp}} \left(\frac{\lambda_4 \pm i\lambda_5}{\sqrt{2}}\right) + \underline{\Delta}_{\bar{K}^0/K^0} \left(\frac{\lambda_6 \pm i\lambda_7}{\sqrt{2}}\right) + \underline{\Delta}_8 \lambda_8 \tag{32}$$

and

$$\sum_{a=0}^{8} \underline{\pi}_a \lambda_a^T = \underline{\Delta}_0 \lambda_0 + \underline{\Delta}_{\pi^{\mp}} \left(\frac{\lambda_1 \mp i\lambda_2}{\sqrt{2}}\right) + \underline{\Delta}_{\pi^0} \lambda_3 + \underline{\Delta}_{K^{\mp}} \left(\frac{\lambda_4 \mp i\lambda_5}{\sqrt{2}}\right) + \underline{\Delta}_{\bar{K}^0/K^0} \left(\frac{\lambda_6 \mp i\lambda_7}{\sqrt{2}}\right) + \underline{\Delta}_8 \lambda_8. \tag{33}$$

with $\underline{\Delta}_0 = \underline{\pi}_0$, $\underline{\Delta}_{\pi^{\mp}} = \frac{\underline{\pi}_1 \mp i\underline{\pi}_2}{\sqrt{2}}$, $\underline{\Delta}_{\pi^0} = \underline{\pi}_3$, $\underline{\Delta}_{K^{\mp}} = \frac{\underline{\pi}_4 \mp i\underline{\pi}_5}{\sqrt{2}}$, $\underline{\Delta}_{\bar{K}^0/K^0} = \frac{\underline{\pi}_6 \mp i\underline{\pi}_7}{\sqrt{2}}$ and $\underline{\Delta}_8 = \underline{\pi}_8$. This rewriting is performed with the scalar mesons in the same way.

Then, Eq. (28) is modified in order to extract three distinct terms [58,59,62,63]. The first concerns the mesons and its expression is

$$S_{Mesons} = \frac{i}{4} Tr \left( S^+ \underline{\Sigma} S^+ \underline{\Sigma} + S^- \underline{\Sigma}^T S^- \underline{\Sigma}^T + \Xi^- \underline{\Sigma}^T \Xi^+ \underline{\Sigma} + \Xi^+ \underline{\Sigma} \Xi^- \underline{\Sigma}^T \right). \tag{34}$$

The second involves the diquarks,

$$S_{Diquarks} = \frac{i}{4} Tr \left( \Xi^- \underline{\Delta}^+ \Xi^- \underline{\Delta}^+ + \Xi^+ \underline{\Delta}^- \Xi^+ \underline{\Delta}^- + S^+ \underline{\Delta}^- S^- \underline{\Delta}^+ + S^- \underline{\Delta}^+ S^+ \underline{\Delta}^- \right). \tag{35}$$

The third reveals a coupling between the mesons and diquarks,

$$S_{Mixed} = \frac{i}{4} Tr \begin{pmatrix} S^+ \underline{\Sigma} \Xi^- \underline{\Delta}^+ + \Xi^- \underline{\Delta}^+ S^+ \underline{\Sigma} + S^+ \underline{\Delta}^- \Xi^+ \underline{\Sigma} + \Xi^+ \underline{\Sigma} S^+ \underline{\Delta}^- \\ + S^- \underline{\Sigma}^T \Xi^+ \underline{\Delta}^- + \Xi^+ \underline{\Delta}^- S^- \underline{\Sigma}^T + S^- \underline{\Delta}^+ \Xi^- \underline{\Sigma}^T + \Xi^- \underline{\Sigma}^T S^- \underline{\Delta}^+ \end{pmatrix}. \tag{36}$$

Appendix C details this term and underlines what mesons and diquarks are coupled in the 2SC phase, with or without the isospin symmetry. This coupling is discussed in Refs. [58,65,66]. However, as in Refs. [59,62], one neglects $\mathcal{S}_{Mixed}$.

## B. The mesons

### 1. The employed method

The pseudoscalar kaons are the "simplest" mesons in this modeling, since they involve $\Xi^{\pm}=0$ in the 2SC regime. Concerning $K^+$, adapting the method detailed in Refs. [58,59,62], one writes

$$\frac{-\partial^2 \mathcal{S}_{Mesons}}{\partial \underline{\Delta}_{K^-} \partial \underline{\Delta}_{K^+}} = -\frac{i}{2} Tr\left(S_u^+ i\gamma_5 S_s^+ i\gamma_5 + S_s^- i\gamma_5 S_u^- i\gamma_5\right). \tag{37}$$

$Tr\left(S_u^+ i\gamma_5 S_s^+ i\gamma_5 + S_s^- i\gamma_5 S_u^- i\gamma_5\right) = 2Tr\left(S_u^+ i\gamma_5 S_s^+ i\gamma_5\right)$, where the 2 in front of the trace is a flavor factor. Moreover, this trace is rewritten as

$$2Tr\left(S_u^+ i\gamma_5 S_s^+ i\gamma_5\right) = 2\sum_{c=r,g,b} Tr\left(S_u^{c+} i\gamma_5 S_s^{c+} i\gamma_5\right) \equiv 2Tr\left(S_{u\bar{s}}\right), \tag{38}$$

with the shorthand notation

$$S_{f\bar{f}'} = \sum_{c=r,g,b} S_f^{c+} i\gamma_5 S_{f'}^{c+} i\gamma_5. \tag{39}$$

Then, the integral over time and space gives an integral over energy and impulsion, via a Fourier Transform. In the Matsubara formalism [81], the integration over energy is replaced by a summation over the fermionic Matsubara frequencies $\omega_n = (2n+1)\pi T$, with $n \in \mathbb{Z}$, i.e.

$$\int d^4x \xrightarrow{\text{Fourier Transform}} \int \frac{d^4p}{(2\pi)^4} \xrightarrow{\text{Matsubara Formalism}} \frac{i}{\beta} \sum_{n=-\infty}^{+\infty} \int \frac{d^3\vec{p}}{(2\pi)^3}. \tag{40}$$

This methodology allows obtaining an expression of a loop function

$$-i\Pi_{u\bar{s}}^{i\gamma_5}\left(iv_m, \vec{k}\right) = -\frac{i}{\beta} \sum_{c=r,g,b} \sum_{n=-\infty}^{+\infty} \int \frac{d^3\vec{p}}{(2\pi)^3} Tr\left[iS_u^{c+}(i\omega_n, \vec{p}) i\gamma_5\, iS_s^{c+}(i\omega_n - iv_m, \vec{p}-\vec{k}) i\gamma_5\right], \tag{41}$$

which is easily comparable to the irreducible polarization function of Refs. [50,53,54]. In this relation, $v_m = 2m\pi T$ is a bosonic Matsubara frequency, with $m \in \mathbb{Z}$, and corresponds to the total energy of the described meson. $\vec{k}$ is its impulsion.

The derivation of the equations is strictly similar for the other kaons. Indeed, with $K^-$, $K^0$ and $\bar{K}^0$, one uses, respectively, $\frac{-\partial^2 \mathcal{S}_{Mesons}}{\partial \underline{\Delta}_{K^+} \partial \underline{\Delta}_{K^-}}$, $\frac{-\partial^2 \mathcal{S}_{Mesons}}{\partial \underline{\Delta}_{\bar{K}^0} \partial \underline{\Delta}_{K^0}}$ and $\frac{-\partial^2 \mathcal{S}_{Mesons}}{\partial \underline{\Delta}_{K^0} \partial \underline{\Delta}_{\bar{K}^0}}$, and one obtains $2Tr(S_{s\bar{u}})$, $2Tr(S_{d\bar{s}})$ and $2Tr(S_{s\bar{d}})$.

Concerning the pion $\pi^+$, the description includes $\Xi^{\pm}$ terms in the 2SC phase. Consequently, $\frac{-\partial^2 \mathcal{S}_{Mesons}}{\partial \underline{\Delta}_{\pi^-} \partial \underline{\Delta}_{\pi^+}}$ gives the trace

$$Tr\left(2S_{u\bar{d}} + \Xi_{ud}^{rg-} i\gamma_5 \Xi_{ud}^{gr+} i\gamma_5 + \Xi_{ud}^{gr-} i\gamma_5 \Xi_{ud}^{rg+} i\gamma_5 + \Xi_{du}^{rg+} i\gamma_5 \Xi_{du}^{gr-} i\gamma_5 + \Xi_{du}^{gr+} i\gamma_5 \Xi_{du}^{rg-} i\gamma_5\right). \tag{42}$$

In the framework of the isospin symmetry,

$$\Xi_{du}^{cc'+} i\gamma_5 \Xi_{du}^{cc'-} i\gamma_5 = \Xi_{ud}^{cc'+} i\gamma_5 \Xi_{ud}^{cc'-} i\gamma_5 \tag{43}$$

and

$$Tr\left(\Xi_{du}^{rg+} i\gamma_5 \Xi_{du}^{gr-} i\gamma_5\right) = Tr\left(\Xi_{ud}^{rg-} i\gamma_5 \Xi_{ud}^{gr+} i\gamma_5\right), \tag{44}$$

so that the trace becomes $2Tr\left(S_{u\bar{d}} + \Xi_{ud}^{rg-} i\gamma_5 \Xi_{ud}^{gr+} i\gamma_5 + \Xi_{ud}^{gr-} i\gamma_5 \Xi_{ud}^{rg+} i\gamma_5\right)$ with obviously $S_{u\bar{d}} \equiv S_{q\bar{q}}$. Finally, one gets the expression

$$2Tr\left(S_{q\bar{q}} + \Xi_q\right) \tag{45}$$

with

$$\Xi_q \equiv \Xi_{ud}^{rg-} i\gamma_5 \Xi_{ud}^{gr+} i\gamma_5 + \Xi_{ud}^{gr-} i\gamma_5 \Xi_{ud}^{rg+} i\gamma_5. \tag{46}$$

The pion $\pi^-$ is described with $\dfrac{-\partial^2 \mathcal{S}_{Mesons}}{\partial \underline{\Delta}_{\pi^+} \partial \underline{\Delta}_{\pi^-}}$. When the isospin symmetry is applied, Eq. (45) is found again, as expected.

Because of the $\pi^0, \eta, \eta'$ coupling, the method described in Refs. [53,54] is employed to model them. The details of the calculations are proposed in Appendix D.

*2. Detailed expression of the loop function*

One considers two red quarks *1* and *2* that interact via a pseudoscalar interaction. The first goes towards the future and the second towards the past. In the 2SC regime and with the isospin symmetry, the general expression of this quark-antiquark loop is

$$Tr\left[S_{f_1}^{r+}(i\omega_n, \vec{p}) \gamma_5 S_{f_2}^{r+}(i\omega_n - i\nu_m, \vec{p}-\vec{k}) \gamma_5\right] = -2\left( \frac{N_1^g}{\prod_{j=1}^{4}(i\omega_n - \lambda_{1,j})} + \frac{N_2^g}{\prod_{j=1}^{4}(i\omega_n - i\nu_m - \lambda_{2,j})} \right.$$

$$\left. + \frac{\left\{N_1^g N_2^g \left[(m_1 - m_2)^2 - (i\nu_m + \mu_1^r - \mu_2^r)^2 + \vec{k}^2 + |\Delta_1|^2 + |\Delta_2|^2\right] + 8\tilde{\mu}_1 \tilde{\mu}_2 |\Delta_1|^2 |\Delta_2|^2 \right\}}{\left[\prod_{j=1}^{4}(i\omega_n - \lambda_{1,j})\right]\left[\prod_{j=1}^{4}(i\omega_n - i\nu_m - \lambda_{2,j})\right]} \right), \tag{47}$$

where $N_1^g = (i\omega_n - \mu_1^g)^2 - E_1^2 - |\Delta_1|^2$, $N_2^g = (i\omega_n - i\nu_m - \mu_2^g)^2 - E_2^2 - |\Delta_2|^2$, $E_1 = \sqrt{\vec{p}^2 + m_1^2}$, $E_2 = \sqrt{(\vec{p}-\vec{k})^2 + m_2^2}$, $\tilde{\mu}_i = \dfrac{\mu_i^r + \mu_i^g}{2}$ and obviously $\mu_1^r - \mu_2^r = \mu_1 - \mu_2 \in \mathbb{R}$. The denominators that appear in this relation are formed of terms like

$$\prod_{j=1}^{4}(x - \lambda_{i,j}) = \left[(x + \mu_i^r + E_i)(x - \mu_i^g - E_i) - |\Delta_i|^2\right]\left[(x + \mu_i^r - E_i)(x - \mu_i^g + E_i) - |\Delta_i|^2\right]. \tag{48}$$

In other words, the $\lambda_{i,j}$ are the four roots of the quartic polynomial visible in Eq. (48). When the flavor of the quark *1* is *u* or *d*, $\Delta_1 \equiv \Delta_{ud}$; otherwise $\Delta_1 = 0$, and so on for the quark *2*. This writing allows Eq. (47) to be valid for all the treated $SU(3)_f$ pseudoscalar mesons. Furthermore, the trace

$$Tr\left[S_{f_1}^{g+}(i\omega_n, \vec{p}) \gamma_5 S_{f_2}^{g+}(i\omega_n - i\nu_m, \vec{p}-\vec{k}) \gamma_5\right] \tag{49}$$

is obtained with the exchange $r \leftrightarrow g$ in Eq. (47), including in the $\lambda_{i,j}$. Since the blue quarks are not coupled in the 2SC phase,

$$Tr\left[S_{f_1}^{b+}(i\omega_n,\vec{p})\gamma_5 S_{f_2}^{b+}(i\omega_n-i\nu_m,\vec{p}-\vec{k})\gamma_5\right] = -2\left(\frac{1}{(i\omega_n+\mu_1^b)^2-E_1^2}\right.$$

$$\left.+\frac{1}{(i\omega_n-i\nu_m+\mu_2^b)^2-E_2^2}+\frac{(m_1-m_2)^2-(i\nu_m+\mu_1-\mu_2)^2+\vec{k}^2}{\left[(i\omega_n+\mu_1^b)^2-E_1^2\right]\left[(i\omega_n-i\nu_m+\mu_2^b)^2-E_2^2\right]}\right). \quad (50)$$

This expression recalls the relations found in Ref. [50]. Indeed, in this NJL Ref., the irreducible polarization function of the pseudoscalar mesons is expressed as

$$-i\Pi_{q_1,\bar{q}_2}^{i\gamma_5}(i\nu_m,\vec{k}) = i\frac{N_c}{8\pi^2}\left\{A(m_1,\mu_1)+A(m_2,\mu_2)+\left[(m_1-m_2)^2-(i\nu_m+\mu_1-\mu_2)^2+\vec{k}^2\right]\right.$$
$$\left.\times B_0(\vec{k},m_1,\mu_1,m_2,\mu_2,\mathrm{Re}(i\nu_m))\right\}, \quad (51)$$

where the $A$ and $B_0$ functions [57] are defined as

$$A(m,\mu) = \frac{16\pi^2}{\beta}\sum_n\int\frac{d^3\vec{p}}{(2\pi)^3}\frac{1}{(i\omega_n+\mu)^2-E^2} \quad (52)$$

and

$$B_0(\vec{k},m_1,\mu_1,m_2,\mu_2,\mathrm{Re}(i\nu_m)) = \frac{16\pi^2}{\beta}\sum_n\int\frac{d^3\vec{p}}{(2\pi)^3}\frac{1}{(i\omega_n+\mu_1)^2-E_1^2}\frac{1}{(i\omega_n-\mathrm{Re}(i\nu_m)+\mu_2)^2-E_2^2}. \quad (53)$$

The Fermi-Dirac distributions $f_{FD}(E_f\mp\mu_f)$ appear when the Matsubara summation is performed in Eqs. (52) and (53) [53,57]. In the PNJL literature [32,34], the only modification is their replacement by $f_\Phi^\pm(E_f\mp\mu_f)$, which correspond to a color averaging of these distributions [48], i.e.

$$f_\Phi^\pm(E_f\mp\mu_f) = \frac{1}{N_c}\sum_{c=r,g,b}f_{FD}(E_f\mp\mu_f^c). \quad (54)$$

The similarities between Eqs. (47), (50) and (51) can be pointed out, notably because of the presence of $(m_1-m_2)^2-(i\nu_m+\mu_1-\mu_2)^2+\vec{k}^2$ in these equations. In addition, with the replacements $i\omega_n-i\nu_m\to-i\omega_n$ and $\vec{p}\to\vec{k}-\vec{p}$ [57], I verify that

$$Tr\left[S_{f_2}^{c-}(i\omega_n,\vec{p})\gamma_5 S_{f_1}^{c-}(i\omega_n-i\nu_m,\vec{p}-\vec{k})\gamma_5\right] = Tr\left[S_{f_1}^{c+}(i\omega_n,\vec{p})\gamma_5 S_{f_2}^{c+}(i\omega_n-i\nu_m,\vec{p}-\vec{k})\gamma_5\right]. \quad (55)$$

This relation is strictly exact in the case of an infinite cutoff or $\vec{k}=\vec{0}$.

Moreover, if the $1$ and $2$ quarks are $u$ or $d$ with the colors $r$ or $g$,

$$Tr\left[\Xi_{ud}^{rg-}(i\omega_n,\vec{p})\gamma_5\Xi_{ud}^{gr+}(i\omega_n-i\nu_m,\vec{p}-\vec{k})\gamma_5\right]$$

$$=\frac{4|\Delta_{ud}|^2\left\{\begin{array}{l}\left[(i\omega_n+\mu_q^r)(i\omega_n-\mu_q^g)-E_1^2-|\Delta_{ud}|^2\right]\\ \times\left[(i\omega_n-i\nu_m+\mu_q^r)(i\omega_n-i\nu_m-\mu_q^g)-E_2^2-|\Delta_{ud}|^2\right]\\ +2\tilde{\mu}_q^2(-E_1^2-E_2^2+\vec{k}^2)\end{array}\right\}}{\left[\prod_{j=1}^4(i\omega_n-\lambda_{1,j})\right]\left[\prod_{j=1}^4(i\omega_n-i\nu_m-\lambda_{2,j})\right]}, \quad (56)$$

with $E_1 = \sqrt{\vec{p}^2 + m_q^2}$ and $E_2 = \sqrt{(\vec{p}-\vec{k})^2 + m_q^2}$. Also, $Tr\left[\Xi_{ud}^{rg+}(i\omega_n, \vec{p})\gamma_5 \Xi_{ud}^{gr-}(i\omega_n - i\nu_m, \vec{p}-\vec{k})\gamma_5\right]$ gives the same result as Eq. (56) and $Tr\left[\Xi_{ud}^{gr-}(i\omega_n, \vec{p})\gamma_5 \Xi_{ud}^{rg+}(i\omega_n - i\nu_m, \vec{p}-\vec{k})\gamma_5\right]$ is found via Eq. (A12).

The equations of the scalar mesons are found with the replacement $m_2 \to -m_2$ in Eqs. (47), (49) and (50), and with slight adaptations in Eq. (56).

*3. The equation to be solved for the mesons*

The $G$ and $K$ terms of the Lagrangian density, Eq. (1), are rewritten as [27,50]

$$G\sum_{a=0}^{8}\left[(\bar{\psi}\tau_a\psi)^2 + (\bar{\psi}i\gamma_5\tau_a\psi)^2\right]$$
$$-K\left\{\det_f\left[\bar{\psi}(1+\gamma_5)\psi\right] + \det_f\left[\bar{\psi}(1-\gamma_5)\psi\right]\right\}$$
$$= \sum_{a=0}^{8}\left[K_{aa}^-(\bar{\psi}\tau_a\psi)^2 + K_{aa}^+(\bar{\psi}i\gamma_5\tau_a\psi)^2\right]$$
$$+K_{30}^-(\bar{\psi}\tau_3\psi)(\bar{\psi}\tau_0\psi) + K_{30}^+(\bar{\psi}i\gamma_5\tau_3\psi)(\bar{\psi}i\gamma_5\tau_0\psi)$$
$$+K_{03}^-(\bar{\psi}\tau_0\psi)(\bar{\psi}\tau_3\psi) + K_{03}^+(\bar{\psi}i\gamma_5\tau_0\psi)(\bar{\psi}i\gamma_5\tau_3\psi) \quad . \tag{57}$$
$$+K_{80}^-(\bar{\psi}\tau_8\psi)(\bar{\psi}\tau_0\psi) + K_{80}^+(\bar{\psi}i\gamma_5\tau_8\psi)(\bar{\psi}i\gamma_5\tau_0\psi)$$
$$+K_{08}^-(\bar{\psi}\tau_0\psi)(\bar{\psi}\tau_8\psi) + K_{08}^+(\bar{\psi}i\gamma_5\tau_0\psi)(\bar{\psi}i\gamma_5\tau_8\psi)$$
$$+K_{83}^-(\bar{\psi}\tau_8\psi)(\bar{\psi}\tau_3\psi) + K_{83}^+(\bar{\psi}i\gamma_5\tau_8\psi)(\bar{\psi}i\gamma_5\tau_3\psi)$$
$$+K_{38}^-(\bar{\psi}\tau_3\psi)(\bar{\psi}\tau_8\psi) + K_{38}^+(\bar{\psi}i\gamma_5\tau_3\psi)(\bar{\psi}i\gamma_5\tau_8\psi)$$

where

$$\begin{aligned}
K_{00}^\pm &= G \pm (\sigma_u + \sigma_d + \sigma_s)K/3, & K_{11}^\pm &= K_{22}^\pm = K_{33}^\pm = G \mp \sigma_s K/2,\\
K_{44}^\pm &= K_{55}^\pm = G \mp \sigma_d K/2, & K_{66}^\pm &= K_{77}^\pm = G \mp \sigma_u K/2,\\
K_{88}^\pm &= G \mp (2\sigma_u + 2\sigma_d - \sigma_s)K/6, & K_{03}^\pm &= K_{30}^\pm = \pm(\sigma_u - \sigma_d)\sqrt{6}K/12,\\
K_{08}^\pm &= K_{80}^\pm = \mp(\sigma_u + \sigma_d - 2\sigma_s)\sqrt{2}K/12, & K_{38}^\pm &= K_{83}^\pm = \mp(\sigma_u - \sigma_d)\sqrt{3}K/6.
\end{aligned} \tag{58}$$

$\sigma_f = \langle\langle\bar{\psi}_f\psi_f\rangle\rangle$ is the chiral condensate of the flavor $f$ quarks, Eq. (21). Finally, the masses $m$ at rest of the pions, kaons and their scalar partners satisfy the relation

$$1 - 4\mathcal{Z}\Pi_{f\bar{f}'}(i\nu_m = m, \vec{k} = \vec{0}) = 0, \tag{59}$$

with $\mathcal{Z} = K_{11}^+$ for $\pi^\pm$, $K_{44}^+$ for $K^\pm$, $K_{66}^+$ for $K^0$ and $\bar{K}^0$, $K_{11}^-$ for $a_0^\pm$, $K_{44}^-$ for $K_0^{*\pm}$, $K_{66}^-$ for $K_0^{*0}$ and $\bar{K}_0^{*0}$. When the meson is stable in the model, $m$ is real. Otherwise, it becomes complex,

$$m = m_{\text{physical}} - i\Gamma/2. \tag{60}$$

$\Gamma$ is the width of the meson [50], which is then a resonance. The structure of the equations to be solved to find the masses of $\eta, \eta', f_0, f_0'$ is not modified in comparison to Refs. [27,50]. However, in the 2SC phase, the calculations described in Appendix D should be taken into account.

### C. The diquarks

*1. Derivation of the equations for the diquarks*

Firstly, the formalism of Refs. [58,59,62] is adapted to model the scalar diquarks $[us]$ with

$$\sum_{A=2,5,7} \frac{-\partial^2 \mathcal{S}_{Diquarks}}{\partial \underline{\Delta}_{us}^A \partial \underline{\Delta}_{us}^{A*}} = -\frac{i}{4} Tr \Big( S_{us}^{rg} + S_{su}^{rg} + S_{us}^{gr} + S_{su}^{gr} \\ + S_{us}^{rb} + S_{su}^{rb} + S_{us}^{br} + S_{su}^{br} + S_{us}^{gb} + S_{su}^{gb} + S_{us}^{bg} + S_{su}^{bg} \Big)$$ (61)

using the shorthand notation

$$S_{ff'}^{cc'} = S_f^{c+} i\gamma_5 S_{f'}^{c'-} i\gamma_5.$$ (62)

In the general case, $Tr\left(S_{f'f}^{c'c}\right) = Tr\left(S_{ff'}^{cc'}\right)$, so that Eq. (61) becomes

$$\sum_{A=2,5,7} \frac{-\partial^2 \mathcal{S}_{Diquarks}}{\partial \underline{\Delta}_{us}^A \partial \underline{\Delta}_{us}^{A*}} = -\frac{i}{2} Tr\left(S_{us}^{rg} + S_{us}^{gr}\right) - \frac{i}{2} Tr\left(S_{us}^{rb} + S_{us}^{br}\right) - \frac{i}{2} Tr\left(S_{us}^{gb} + S_{us}^{bg}\right).$$ (63)

These three traces describe, respectively, $[us]^{rg}$, $[us]^{rb}$ and $[us]^{gb}$. The methodology described in Sect. III.B.1 is then used again. With $[us]^{rg}$, one gets the expression

$$-i\Pi_{us}^{rg\,(i\gamma_5)}\left(iv_m, \vec{k}\right) = -4 \frac{i}{\beta} \sum_{n=-\infty}^{+\infty} \int \frac{d^3 \vec{p}}{(2\pi)^3} Tr\Big[ iS_u^{r+}\left(i\omega_n, \vec{p}\right) i\gamma_5 \, iS_s^{g-}\left(i\omega_n - iv_m, \vec{p} - \vec{k}\right) i\gamma_5 \\ + iS_u^{g+}\left(i\omega_n, \vec{p}\right) i\gamma_5 \, iS_s^{r-}\left(i\omega_n - iv_m, \vec{p} - \vec{k}\right) i\gamma_5 \Big]$$ (64)

and so on for $[us]^{rb}$ and $[us]^{gb}$. The antidiquarks $\overline{[us]}$ are described with $\sum_{A=2,5,7} \frac{-\partial^2 \mathcal{S}_{Diquarks}}{\partial \underline{\Delta}_{us}^{A*} \partial \underline{\Delta}_{us}^{A}}$ in the same way. In Eqs. (62) and (64), one only needs to exchange $+$ and $-$ in the $S_f^{c\pm}$. Concerning the scalar diquarks $[ds]$ and antidiquarks $\overline{[ds]}$, the calculations are identical with the replacement $u \to d$.

The modeling of the diquarks $[ud]^{rb}$, $[ud]^{gb}$ and the corresponding antidiquarks are performed in the same manner. However, the scalar diquarks $[ud]^{rg}$ and antidiquarks $\overline{[ud]}^{rg}$ can be coupled [66]. This leads to additional terms in their description. Indeed, $\frac{-\partial^2 \mathcal{S}_{Diquarks}}{\partial \underline{\Delta}_{ud}^2 \partial \underline{\Delta}_{ud}^2}$ and $\frac{-\partial^2 \mathcal{S}_{Diquarks}}{\partial \underline{\Delta}_{ud}^{2*} \partial \underline{\Delta}_{ud}^{2*}}$ are not null in the 2SC phase. Consequently, one considers the matrix

$$\tilde{D} = \begin{bmatrix} A = \dfrac{-\partial^2 \mathcal{S}_{Diquarks}}{\partial \underline{\Delta}_{ud}^2 \partial \underline{\Delta}_{ud}^{2*}} & B = \dfrac{-\partial^2 \mathcal{S}_{Diquarks}}{\partial \underline{\Delta}_{ud}^2 \partial \underline{\Delta}_{ud}^2} \\ C = \dfrac{-\partial^2 \mathcal{S}_{Diquarks}}{\partial \underline{\Delta}_{ud}^{2*} \partial \underline{\Delta}_{ud}^{2*}} & \overline{A} = \dfrac{-\partial^2 \mathcal{S}_{Diquarks}}{\partial \underline{\Delta}_{ud}^{2*} \partial \underline{\Delta}_{ud}^2} \end{bmatrix},$$ (65)

with

$$A = -\frac{i}{2} Tr\left(S_u^{r+} i\gamma_5 S_d^{g-} i\gamma_5 + S_u^{g+} i\gamma_5 S_d^{r-} i\gamma_5\right), \quad \overline{A} = -\frac{i}{2} Tr\left(S_u^{r-} i\gamma_5 S_d^{g+} i\gamma_5 + S_u^{g-} i\gamma_5 S_d^{r+} i\gamma_5\right) \\ B = \frac{i}{2} Tr\left(\Xi_{ud}^{rg+} i\gamma_5 \Xi_{ud}^{rg+} i\gamma_5 + \Xi_{ud}^{gr+} i\gamma_5 \Xi_{ud}^{gr+} i\gamma_5\right), \quad C = \frac{i}{2} Tr\left(\Xi_{ud}^{rg-} i\gamma_5 \Xi_{ud}^{rg-} i\gamma_5 + \Xi_{ud}^{gr-} i\gamma_5 \Xi_{ud}^{gr-} i\gamma_5\right)$$ (66)

This matrix has two eigenvalues

$$L^{\pm} = \frac{1}{2}\left[A + \overline{A} \pm \sqrt{(A - \overline{A})^2 + 4BC}\right],$$ (67)

where $L^+$ corresponds to $[ud]^{rg}$ and $L^-$ to $\overline{[ud]}^{rg}$. Furthermore, I have established that $B = C\big|_{\Delta_{ud} \to \Delta_{ud}^*}$ when the isospin symmetry is applied.

## 2. Expression of the loop function and equation to be solved

The loop functions of the scalar diquarks are formed of traces whose expression is, e.g.,

$$
\begin{aligned}
Tr\left[ S_{f_1}^{r+}(i\omega_n,\vec{p})\gamma_5 S_{f_2}^{g-}(i\omega_n-i\nu_m,\vec{p}-\vec{k})\gamma_5 \right] \\
= -2 \left( \frac{\dfrac{N_1^g}{\prod_{j=1}^{4}(i\omega_n-\lambda_{1,j})} + \dfrac{\bar{N}_2^r}{\prod_{j=1}^{4}(i\omega_n-i\nu_m-\lambda_{2,j})}}{} \right. \\
\left. + \frac{\left\{ N_1^g \bar{N}_2^r \left[(m_1-m_2)^2 - (i\nu_m+\mu_1^r+\mu_2^g)^2 + \vec{k}^2 + |\Delta_1|^2 + |\Delta_2|^2 \right] - 8\tilde{\mu}_1 \tilde{\mu}_2 |\Delta_1|^2 |\Delta_2|^2 \right\}}{\left[\prod_{j=1}^{4}(i\omega_n-\lambda_{1,j})\right]\left[\prod_{j=1}^{4}(i\omega_n-i\nu_m-\lambda_{2,j})\right]} \right. \\
\left. + \frac{-4N_1^g \tilde{\mu}_2 |\Delta_2|^2 (i\omega_n+\mu_1^r+\tilde{\mu}_2) + 4\bar{N}_2^r \tilde{\mu}_1 |\Delta_1|^2 (i\omega_n-i\nu_m-\mu_2^g-\tilde{\mu}_1)}{\left[\prod_{j=1}^{4}(i\omega_n-\lambda_{1,j})\right]\left[\prod_{j=1}^{4}(i\omega_n-i\nu_m-\lambda_{2,j})\right]} \right),
\end{aligned} \quad (68)
$$

with $\bar{N}_2^r = (i\omega_n - i\nu_m + \mu_2^r)^2 - E_2^2 - |\Delta_2|^2$. The other notations are the same as in Eq. (47), including the $\lambda_{i,j}$. This Eq. (68) has been established with the isospin symmetry. It can be compared to the loop function used in the (P)NJL literature [53,54], i.e.

$$
\begin{aligned}
-i\Pi_{q_1,q_2}(i\nu_m,\vec{k}) = \frac{i}{\pi^2}\Big\{ A(m_1,\mu_1) + A(m_2,-\mu_2) + \left[(m_1-m_2)^2 - (i\nu_m+\mu_1+\mu_2)^2 + \vec{k}^2\right] \\
\times B_0(\vec{k},m_1,\mu_1,m_2,-\mu_2,\mathrm{Re}(i\nu_m)) \Big\}
\end{aligned} \quad (69)
$$

where the $A$ and $B_0$ functions have been defined Eqs. (52) and (53).

Moreover, I found

$$
Tr\left[ \Xi_{ud}^{rg-}(i\omega_n,\vec{p})\gamma_5 \Xi_{ud}^{rg-}(i\omega_n-i\nu_m,\vec{p}-\vec{k})\gamma_5 \right]
$$

$$
= \frac{4\Delta_{ud}^2 \left\{ \begin{array}{l} \left[(i\omega_n+\mu_q^r)(i\omega_n-\mu_q^g) - E_1^2 - |\Delta_{ud}|^2\right] \\ \times \left[(i\omega_n-i\nu_m+\mu_q^r)(i\omega_n-i\nu_m-\mu_q^g) - E_2^2 - |\Delta_{ud}|^2\right] \\ -2\tilde{\mu}_q^2\left(-E_1^2-E_2^2+\vec{k}^2\right) \end{array} \right\}}{\left[\prod_{j=1}^{4}(i\omega_n-\lambda_{1,j})\right]\left[\prod_{j=1}^{4}(i\omega_n-i\nu_m-\lambda_{2,j})\right]}. \quad (70)
$$

This Eq. is comparable with Eq. (56) and uses the same notations. Moreover, $Tr\left[\Xi_{ud}^{gr-}(i\omega_n,\vec{p})\gamma_5 \Xi_{ud}^{gr-}(i\omega_n-i\nu_m,\vec{p}-\vec{k})\gamma_5\right]$ is obtained with Eq. (A12). As with the scalar mesons, the equations of the pseudoscalar diquarks are found with slight modifications of these relations.

The mass $m$ at rest of a diquark is obtained with [53,54]

$$
1 - 2G_{DIQ}\Pi_{ff'}(i\nu_m = m, \vec{k}=\vec{0}) = 0. \quad (71)
$$

As with Eq. (59), $m$ is real when the diquark is stable, and complex otherwise, Eq. (60) [61]. Also, $\Pi_{ff'}$ is replaced by the expressions found with $L^+$ or $L^-$ in the cases of, respectively, $[ud]^{rg}$ or $\overline{[ud]}^{rg}$. In this paper, a diagonalization of the matrix $\tilde{D}$ is performed to model them, whereas Refs. [58,59,61,62] use a determinant. However, both approaches are equivalent: since $\det(\tilde{D}) = L^+ L^-$,

$$\det\left(1 - 2G_{DIQ}\tilde{D}\right) = \left(1 - 2G_{DIQ}A\right)\left(1 - 2G_{DIQ}\overline{A}\right) - 4G_{DIQ}^2 BC = \left(1 - 2G_{DIQ}L^+\right)\left(1 - 2G_{DIQ}L^-\right). \quad (72)$$

## IV. RESULTS

### A. The mesons

In order to analyze the data obtained with the P1 parameter set in the color superconducting regime, FIG 6 to FIG 13, one proceeds as follow. Firstly, the pseudoscalar and scalar mesons are studied at a fixed temperature. Then, the pseudoscalar mesons are described in the $\rho_B, T$ and $\mu_q, T$ planes. These results have been found with the $\Delta$ NJL model. The differences observed in the $\Delta(\mu)$PNJL descriptions are then presented in the end of this Sect. In comparison with the literature, the main novelties carried out by this work are the inclusion of the strange quarks, the description in the $\rho_B, T$ plane and the three-color $\Delta(\mu)$PNJL modeling. In addition, $\mu_q, T$ calculations of the mesons in the 2SC regime stay rare in the NJL literature, except for Ref. [66].

The FIG 6 and FIG 7 describe the masses of, respectively, the pseudoscalar and scalar mesons according to the baryonic density, at a low temperature. A finality of these graphs is to discuss the influence of certain factors on the masses. One of them is the version of the employed model. In the description of the NJL mesons made in Ref. [50] and then in Refs. [53,54], the complex function $B_0$, Eq. (53), is used to model a two-fermion loop [57]. This function neglects $\text{Im}(iv_m)$, where $iv_m$ is associated with the meson mass, Eq. (59). In FIG 6 and FIG 7, this description corresponds to the NJL curves. In order to investigate the effect of this approximation, I have developed an alternative version of $B_0$ that includes this imaginary part in the treatment. The associated curves are labeled as mNJL. Upon numerical considerations, the calculations with mNJL appears to be more rapid than NJL when the meson mass is complex. Moreover, the NJL and mNJL curves coincide when this mass is real.

Concerning the pseudoscalar mesons at low temperatures, only $K^-$ becomes a resonance, when $\rho_B > 4\rho_0$. FIG 6(b) and FIG 6(d) reveal that the mentioned approximation leads to underestimate the width $\Gamma$ and the mass of this kaon. This remark is generalizable to other mesons, and notably $\eta'$. At low temperatures, $\Gamma_{\eta'}$ strongly diverges in the mNJL description, in such a way that its equations cannot converge. In other words, $\eta'$ has been studied in the NJL model [50,53,54] thanks to this approximation. However, Ref. [50] has mentioned its skepticism about the behavior of this meson in this model, and has attributed it to the lack of confinement. In addition, the equations of $\eta'$ also diverge at high temperatures [54], i.e. when $\Gamma_{\eta'}$ falls to zero, whatever the version of the model. As a consequence, this meson is not treated in this paper.

Nevertheless, this feature does not appear with $f_0'$, FIG 7(a). Concerning the scalar mesons, at low temperatures, $m_{f_0'}$ and $m_{K_0^{*-}}$ are always complex, unlike $m_{f_0}$ that stays real. In addition, $m_{a_0}$ and $m_{K_0^{*+}}$ are complex at low densities, but become real when the density is growing [53]. These behaviors easily explain the mutual evolution of the NJL and mNJL curves in FIG 7.

Moreover, in FIG 6 and FIG 7, the curves labeled as $\Delta = 0$ refer to the data found with the $\Delta$ NJL equations described upstream, but when the gap $\Delta_{ud}$ is fixed to zero. These curves always coincide with the mNJL ones. So, the modeling described in this paper allows finding again the NJL results.

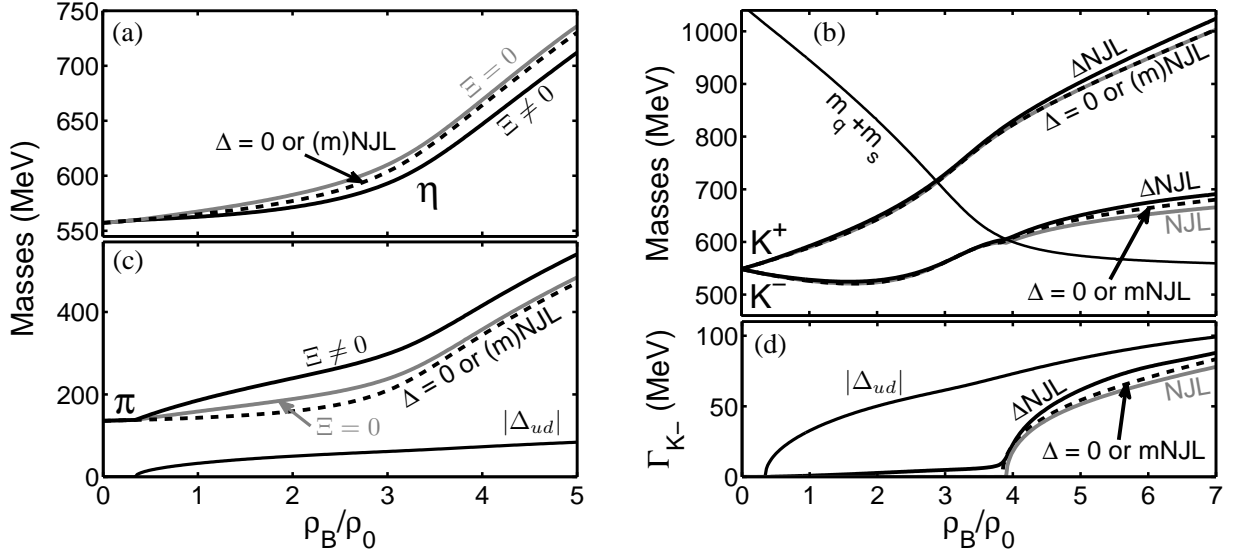

**FIG 6.** Masses of the pseudoscalar mesons (a) $\eta$, (b) $K^{\pm}$, (c) $\pi$ and (d) widths of $K^-$ according to the baryonic density $\rho_B$, at $T = 10$ MeV.

Then, the effects of the 2SC phase on the masses are investigated. Firstly, $\Delta_{ud}$ intervenes in the propagators of the light quarks whose color is red or green, as observed in the previous Sects. and in appendix A. This globally induces an increase in the masses of the mesons formed of $q$ or $\bar{q}$ or both. This contribution appears consistent with the observations of Ref. [48], which indicates that $\Delta_{ud}$ leads to an increase in $m_q$, especially when $\rho_B \approx 3\rho_0$ and at low $T$. Secondly, $\Delta_{ud}$ also intervenes in the $\Xi_q$ term, Eq. (46), which concerns $\pi$, $\eta$, $a_0$, $f_0$ and $f_0'$. In FIG 6(a) and FIG 6(c), the influence of these two contributions is detailed via the curves labeled as $\Xi = 0$ (first contribution only) and $\Xi \neq 0$ (both contributions). The $\Xi_q$ term leads to an increase in $m_\pi$ and a decrease in $m_\eta$, in agreement with the signs placed in front of the $\Xi_q$ in Eqs. (45) and (D5). In absolute values, its influence becomes stronger when $\rho_B$ is growing, i.e. when $\Delta_{ud}$ increases. At $\rho_B = 5\rho_0$, $\Xi_q$ leads to $+58$ MeV for $m_\pi$, vs. $-24$ MeV for $m_\eta$. These gaps are non-negligible, notably for light particles like the pions.

In FIG 6(c), FIG 6(d) and FIG 7, the curves labeled as $\Delta$ NJL include the two contributions evoked in the previous paragraph, if they exist. The observations concerning $\pi$ and $\eta$ are extensible to $a_0$, $f_0$ and $f_0'$. This latter seems to be weakly affected by $\Delta_{ud}$. Moreover, FIG 6(d) reveals that the 2SC phase also intervenes in the evolution of $\Gamma_{K^-}$. With (m)NJL or when $\Delta_{ud}$ is fixed to zero, $\Gamma_{K^-}$ is strictly null when $m_{K^-} < m_q + m_s$. In contrast, the $\Delta$ NJL curve shows that $\Gamma_{K^-}$ is weak but non-zero in this regime if $\Delta_{ud} \neq 0$. Consequently, the transition at $m_{K^-} = m_q + m_s$ is less marked in the 2SC phase. In addition, when $m_{K^-} > m_q + m_s$, $\Gamma_{K^-}$ is stronger with $\Delta$ NJL than with the other versions of the model presented in the graph.

Also, $\pi$ and $f_0$ (also named $\sigma$) becomes degenerate at high densities in the NJL model [22,26]. This observation is still valid in a color superconducting regime, in agreement with Refs. [58,61,62]. A similar behavior is found for $\eta$ with $a_0$, $K^+$ with $K_0^{*+}$ and $K^-$ with $K_0^{*-}$ [53], FIG 6 and FIG 7. Whatever the version of the model, this is due to the restoration of the chiral symmetry at high densities. Indeed, it induces a strong reduction in $\sigma_u, \sigma_d$ that appears in Eq. (58), and in $m_q$ visible in Eqs. (47), (50) and (56).

Moreover, the strength of the meson/diquark coupling is related to the quark mass [58], which is strongly reduced when the chiral symmetry is restored. This constitutes an argument to neglect this coupling, especially at the chiral limit [66]. In addition, Ref. [58] indicates that this coupling weakly affects the coupled mesons and diquarks. This also induces a low width of $f_0$. These observations are strongly in contradiction with Ref. [65] that observes, at high chemical potentials, the loss of the $\pi, f_0$ degeneracy and a non-negligible modification of the masses of the coupled particles. A feature of the two-color description performed in this Ref. is the absence of the first order chiral phase transition. This may explain this disagreement.

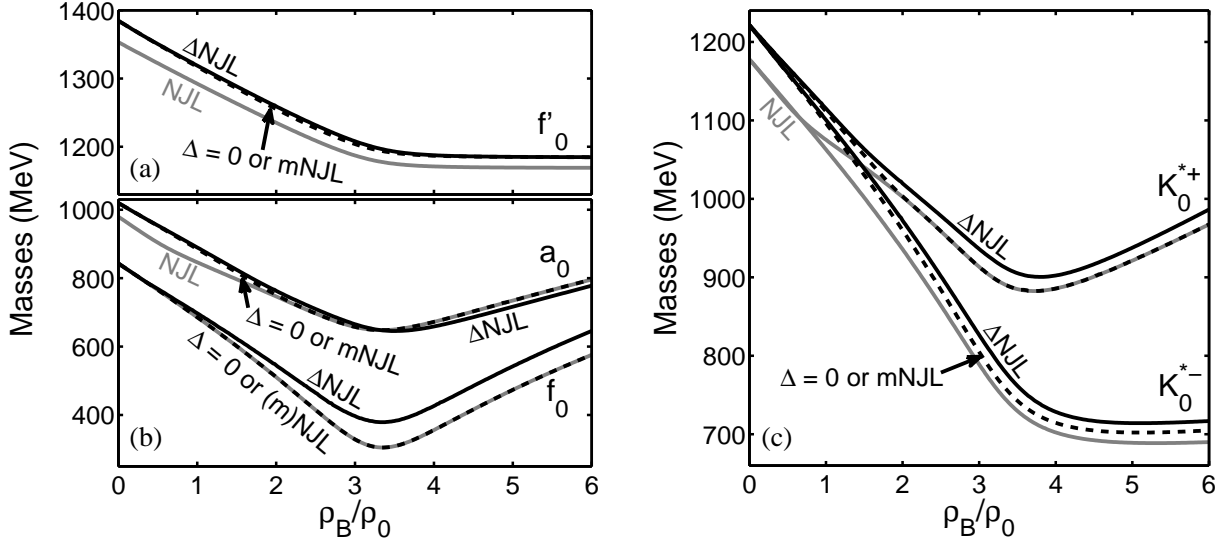

**FIG 7.** Masses of the scalar mesons (a) $f_0'$, (b) $a_0$, $f_0$ and (c) $K_0^{*\pm}$ according to $\rho_B$, at $T = 10$ MeV.

In FIG 8 to FIG 11, the mass calculations are extended to the $\rho_B, T$ plane, for each pseudoscalar meson mentioned upstream. In addition, as in FIG 4, the found data are employed to draw the evolution of these masses in the $\mu_q, T$ plane. At finite temperatures and null density, the NJL pseudoscalar mesons have two distinct regimes: stable at low $T$, resonant/unstable otherwise. This notion of instability refers to the possible disintegration of the meson into a quark-antiquark pair. The stable/unstable transition occurs at the Mott temperature, which depends on the treated meson. After this temperature, the meson mass $m$ becomes significantly complex, and $\text{Re}(m)$ becomes greater than the mass of the quark-antiquark pair that composes the meson. These two features constitute criteria to investigate the limit of the stable regime, for each meson. In FIG 8 to FIG 11, they are materialized by, respectively, a thin and a thick gray curve. These curves coincide in a great part of the $\rho_B, T$ plane.

However, except for $K^-$, a divergence of the $\text{Im}(m) \approx 0$ curve is observed at low temperatures and high densities. In other words, in these conditions, $\pi$, $\eta$ and $K^+$ stay stable whatever the density. This behavior has been previously found in papers that do not consider the color superconductivity [53,54]. Also, in Refs [61,66], $\Gamma_\pi = 0$ at null temperature whatever the chemical potential, including in the 2SC phase. At high $\rho_B$, Pauli exclusion principle (and by extension Pauli blocking) acts on the $u$ or $d$ quarks contained in $\pi$, $\eta$ and $K^+$. This explains the increase in the masses of these mesons at growing $\rho_B$ [26], as well as their observed stability [61]. In contrast, since $K^-$ is made of $s\bar{q}$, its behavior according to $\rho_B$ is different [82-84]: it becomes a resonance at null $T$ when $\rho_B > 4\rho_0$.

Moreover, the modifications of the Fermi-Dirac distributions due to the temperature lead to weaken Pauli blocking. Therefore, $\pi$, $\eta$ and $K^+$ cannot stay stable at high $\rho_B$ when $T$ is strong enough.

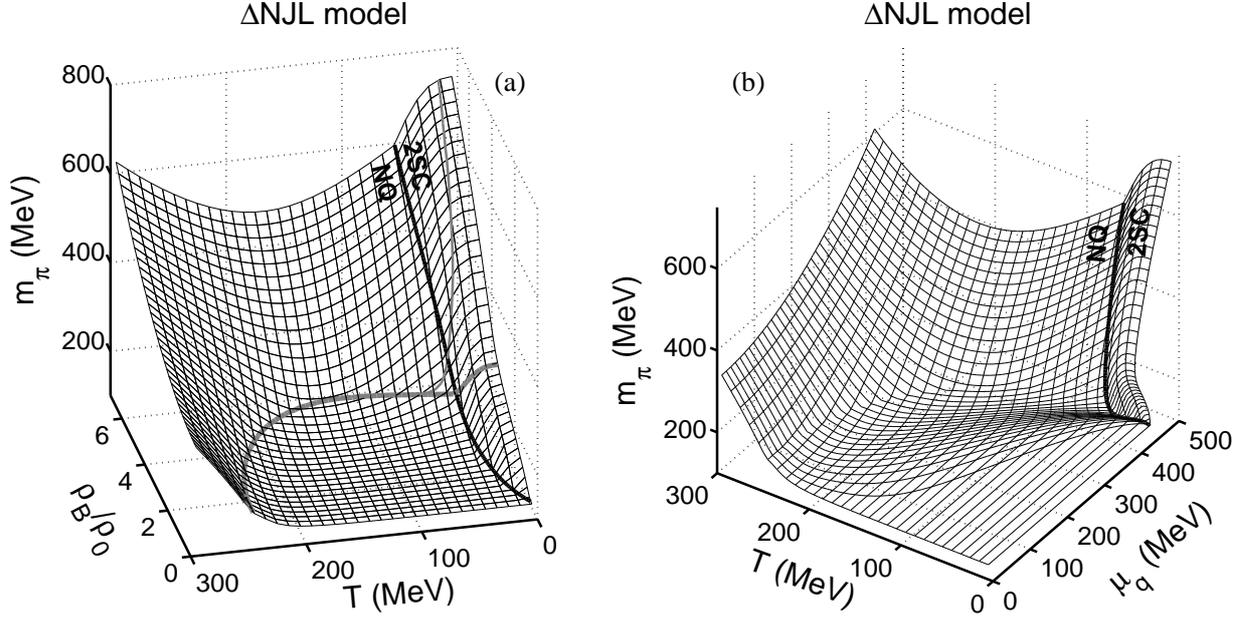

**FIG 8.** Mass of $\pi$ in the (a) $\rho_B,T$ and (b) $\mu_q,T$ planes. The thick gray curve indicates when $m_\pi = 2m_q$ and the thin gray curve is the limit of the zone in which $\text{Im}(m_\pi) \approx 0$.

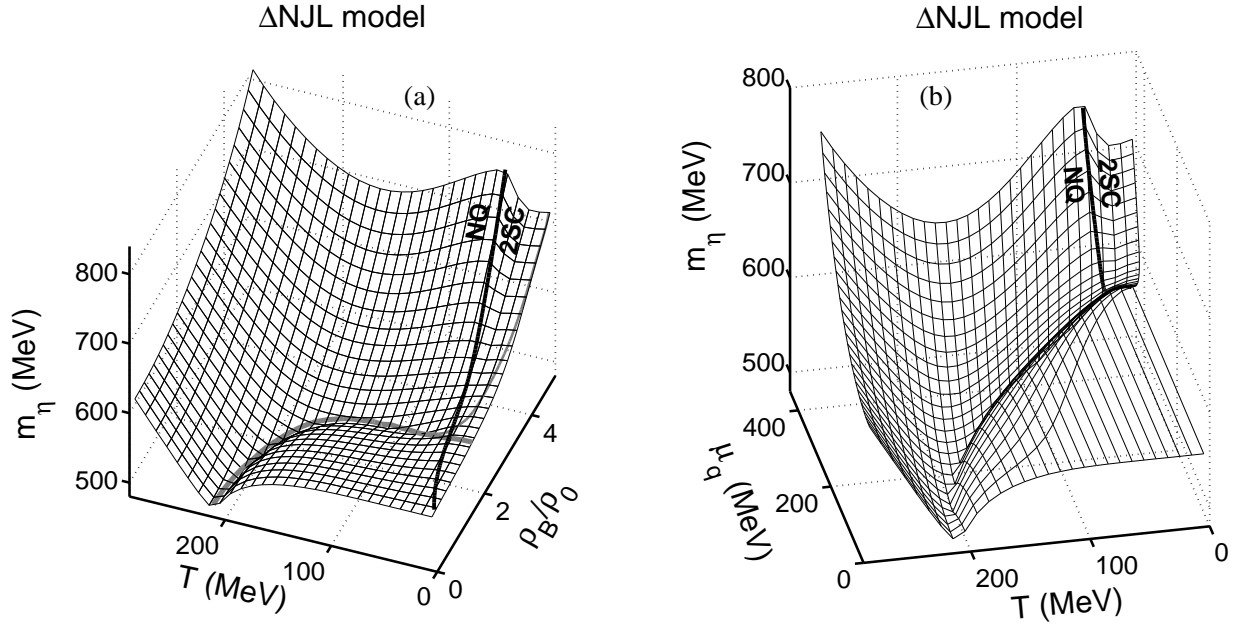

**FIG 9.** Mass of $\eta$ in the (a) $\rho_B,T$ and (b) $\mu_q,T$ planes. The thick gray curve shows when $m_\eta = 2m_q$.

Concerning the influence of the color superconductivity on the masses, the results found in the $\rho_B,T$ plane go on the sense of FIG 6. In FIG 8(a), the mass of $\pi$ is strongly augmented by the 2SC phase, especially at very low $T$ and high $\rho_B$, i.e. when $|\Delta_{ud}|$ reaches the strongest values, FIG 3(b). The effect is reversed for $\eta$ and leads to a notable decreasing in $m_\eta$ in the 2SC regime, FIG 9(a). In contrast, the modifications are modest for $K^+$ and $K^-$, since these mesons are constituted of a strange quark or antiquark, FIG 10 and FIG 11.

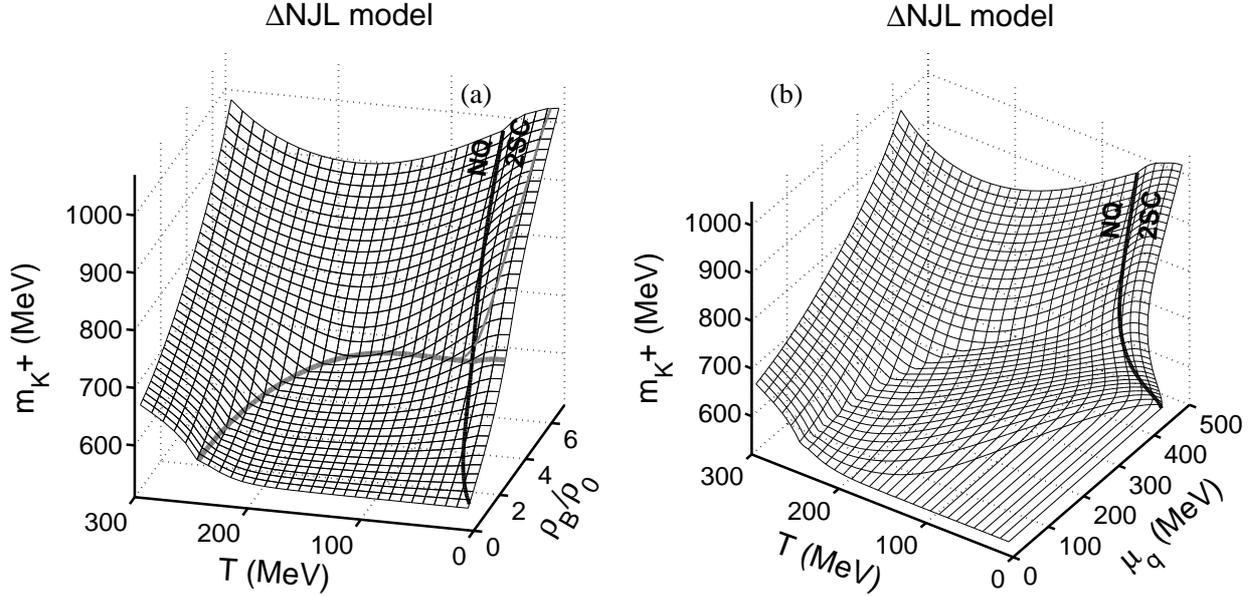

**FIG 10.** Mass of $K^+$ according to (a) $\rho_B, T$ and (b) $\mu_q, T$. The constraint $m_{K^+} = m_q + m_s$ is indicated by the thick gray curve.

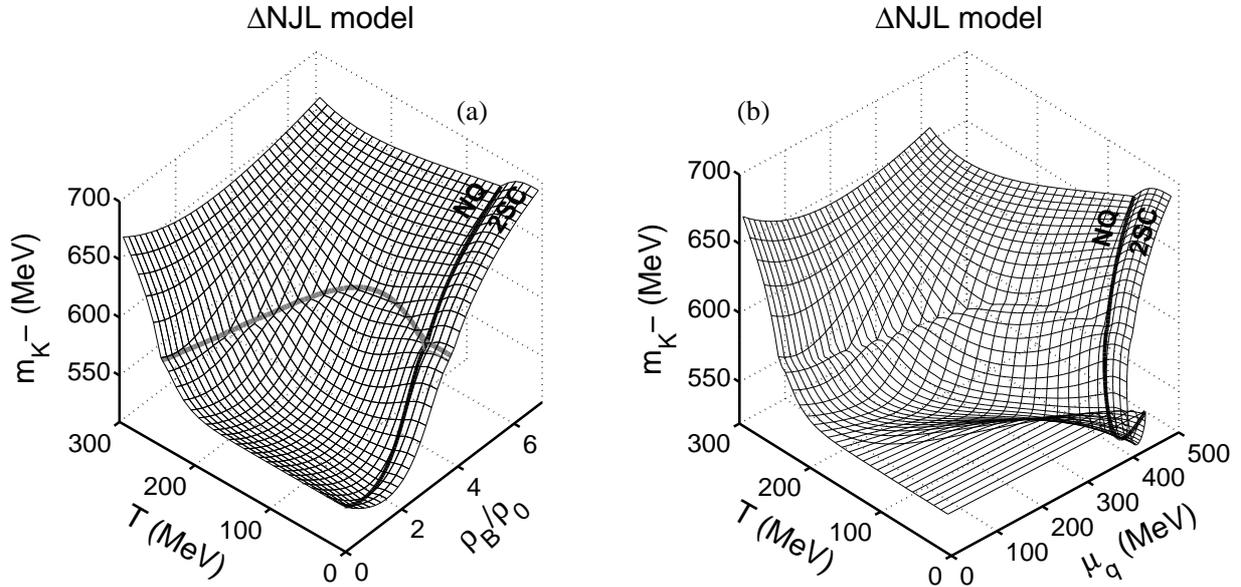

**FIG 11.** Mass of $K^-$ according to (a) $\rho_B, T$ and (b) $\mu_q, T$. The constraint $m_{K^-} = m_q + m_s$ is visible via the thick gray curve.

In the $\mu_q, T$ plane, the four associated graphs allow seeing the strong deformations of the plotted surfaces when the first order chiral phase transition occurs, i.e. at the critical chemical potential $\mu_c \approx 395$ MeV when $T$ is low. These distortions are roughly comparable for $\pi$, $\eta$ and $K^+$. In contrast, the structure visible for $K^-$ has a very different aspect. Indeed, the isotemperature curves draw curls near the chiral phase transition. This behavior is related to the non-monotonic variations of $m_{K^-}$ according to $T$, FIG 6(b), in contradiction with the evolution of the masses of $\pi$, $\eta$ and $K^+$.

In FIG 8(b), the lowest isotemperature curve qualitatively recalls the $m_\pi$ graphs of Refs. [58,61,62]. Indeed, except for the metastable and unstable states near the first order chiral phase transition, $m_\pi$ is constant before $\mu_c$, and grows when $\mu_q > \mu_c$. Also, the results found in this graph at finite $T$ are fully compatible with Ref. [66], even if the increase in $m_\pi$ due to the color superconductivity seems to be lower in this Ref. More generally, at modest $T$, the four $\mu_q, T$ graphs reveal the non-negligible influence of the 2SC phase on the studied mesons, at the level of the chiral phase transition and in the chirally restored regime. So, the inclusion of the color superconductivity in the meson modeling appears relevant to investigate their behavior near this phase transition. In contrast, when the chiral symmetry is broken, i.e. at low $\mu_q$ and $T$, the color superconductivity has no influence on the results.

Moreover, a comparison between NJL and PNJL mesons has been performed in Refs. [32,34,54]. Such works have reported that the Polyakov loop induces a shifting of the mass curves towards higher $T$. At vanishing densities, a modification of their slopes is also visible near the pseudo-critical temperature of the chiral crossover [34]. In contrast, studies of the mesons with $\mu$PNJL are not available in the literature. However, the NJL and $\mu$PNJL models give comparable quark masses. So, this resemblance may be extended to the mesons. FIG 12 confirms this supposition when the color superconductivity is included: FIG 8(a) and FIG 12(a) are very similar. As with the quarks, FIG 3(a) and FIG 5(b), the main differences are found at very low $\rho_B$ and at $T \approx 150$ MeV. Concerning the $\Delta$PNJL curve, FIG 12(b), the mentioned shifting towards higher $T$ is verified. Furthermore, the behavior of the masses at finite temperatures and null density in FIG 8(a) and FIG 12(b) is fully in agreement with (P)NJL publications [32,34,50,54,57]. All the observations performed in this paragraph concern the pions, but are also valid for the other mesons.

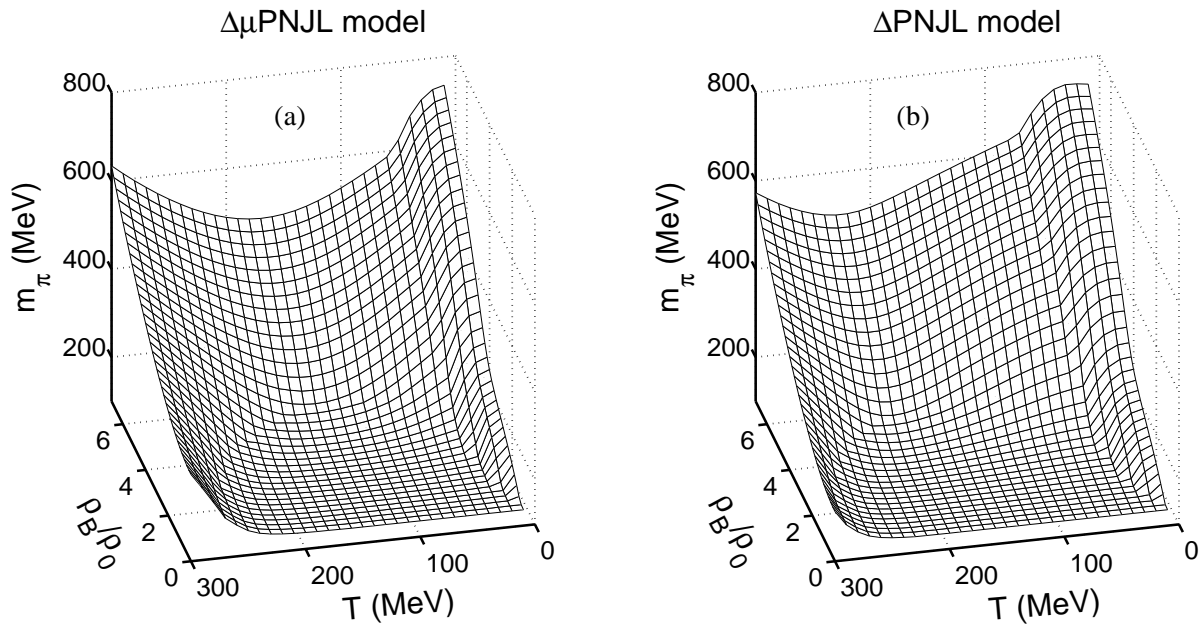

**FIG 12.** Mass of $\pi$ according to $\rho_B$ and $T$, in the (a) $\Delta\mu$ PNJL and (b) $\Delta$ PNJL models.

In FIG 13, the shifting due to the Polyakov loop is also observed with the NQ/2SC frontier, the limit of the unstable zone of the first order chiral phase transition ($\partial^2\Omega/\partial\sigma_q^2 < 0$ criterion), and the

$m_\pi = 2m_q$ curve. Nevertheless, these three curves are not shifted in the same way. In the $\Delta$ PNJL model, $T_{CEP}$ is more affected by the Polyakov loop than the $m_\pi = 2m_q$ curve. These observations are reversed with $\Delta\mu$ PNJL. Moreover, the $\text{Im}(m_\pi) \approx 0$ curve respects the shifting effect of the Polyakov loop in the whole NQ phase, but not in the 2SC regime, FIG 13(b) and FIG 13(c). As in FIG 6(d), the width of the meson starts to be non-null in the 2SC phase at low $\rho_B$, even if it stays very weak. This behavior is rather negligible in the $\Delta$ NJL model, but becomes important with the inclusion of the Polyakov loop. In the $\Delta\mu$ PNJL description, this leads to a deformation of the $\text{Im}(m_\pi) \approx 0$ curve. In the $\Delta$ PNJL model, this effect is stronger, since $m_\pi$ is complex is a great part of the 2SC zone represented in the graph.

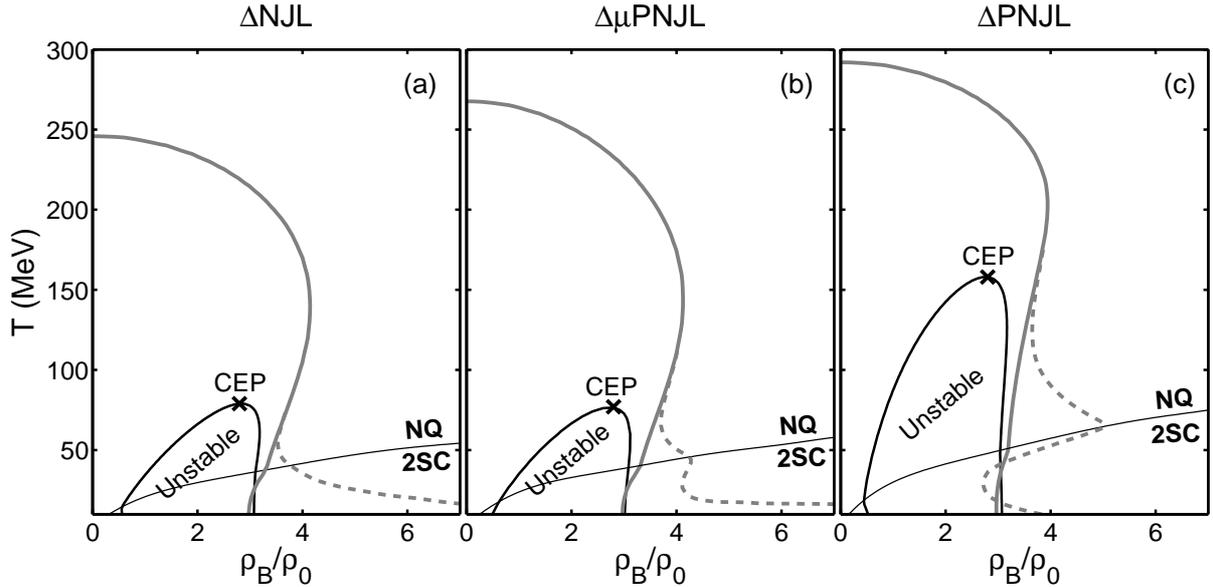

**FIG 13.** Behavior of the pions according to $\rho_B$ and $T$, in the (a) $\Delta$ NJL, (b) $\Delta\mu$ PNJL and (c) $\Delta$ PNJL models. The solid and the dashed gray curves indicate, respectively, when $m_\pi = 2m_q$ and the limit of $\text{Im}(m_\pi) \approx 0$. Both coincide at high $T$. The unstable zone refers to the condition $\partial^2\Omega/\partial\sigma_q^2 < 0$. CEP is the critical endpoint, which delimits the first order chiral phase transition at low $T$ and the crossover at high $T$.

### B. The diquarks

#### 1. $\Delta$ NJL results

The obtained results are presented in the same fashion as in the previous Sect. with the mesons. Nevertheless, important differences should be underlined. A meson is described as a summation on three states $r\bar{r}$, $g\bar{g}$ and $b\bar{b}$, Eq. (39). In contrast, the treated diquarks are antisymmetric in flavor and color. Therefore, three distinct scalar diquarks $[ud]$ are considered: $[ud]^{rg}$, $[ud]^{rb}$ and $[ud]^{gb}$, and so on for the other modeled diquarks and antidiquarks. There is no mixture between them in the performed description [58,59,61,62], except for the coupling visible in Eq. (65). More precisely, the terms found with Eq. (35) are formed of $\Delta_{ff'}^{A}\Delta_{ff'}^{A'*}$, $\Delta_{ff'}^{A}\Delta_{ff'}^{A'}$, $\Delta_{ff'}^{A*}\Delta_{ff'}^{A'*}$ or $\Delta_{ff'}^{A*}\Delta_{ff'}^{A'}$ with necessarily $A = A'$. As a consequence, the three scalar (or pseudoscalar) diquarks $[ff']^{rg}$, $[ff']^{rb}$ and $[ff']^{gb}$ (with $ff' = ud$, $us$ or $ds$) are studied independently, in the NQ and 2SC phases. However, Refs. [63,64] recall that this affirmation cannot be extended to the CFL phase, for which couplings appear. In comparison with the three-color approach of Refs. [58-62], the main novelties of the presented results are the extension at finite $T$ of the calculations, and the use of three flavors.

In the NJL description, the quark colors are unspecified, since they act in the same way. Therefore, the three diquarks $[ud]$ are degenerate, FIG 14(a) and FIG 15(a). This feature is also observed for the same reason with the mNJL and $\Delta = 0$ curves, which strictly coincide. However, a gap is found between NJL and mNJL when $m_{[ud]}$ becomes complex, which corresponds to $\mathrm{Re}(m_{[ud]}) > 2m_q$. The approximation performed in the NJL function $B_0$ [57] induces an underestimation of the diquark masses, but not necessarily of their widths.

When $\Delta_{ud} \neq 0$, the propagators of the $q$ quarks whose color is red or green are modified. Consequently, a partial lifting of degeneracy occurs between $[ud]^{rg}$ and the two other diquarks $[ud]$. In the $\Delta$ NJL description, the red and green $q$ quarks are affected in the same manner by the color superconductivity, which explains the degeneracy of $[ud]^{rb}$ and $[ud]^{gb}$. In addition, the equations of $[ud]^{rg}$ also include $\Xi_{ud}$ terms, Eqs. (66) and (67). They lead to a strong increase in its mass, as attested by the two curves labeled as "$rg\ \Xi = 0$" and "$rg\ \Xi \neq 0$" in FIG 14(a). At $\rho_B = 7\rho_0$, the gap between these curves slightly exceeds 53 MeV. At the same density, the difference between the $\Delta = 0$ and $rg\ \Xi \neq 0$ curves is close to 94 MeV. So, the influence of the 2SC phase on $[ud]^{rg}$ is important and is mainly due to the contribution of the $\Xi_{ud}$.

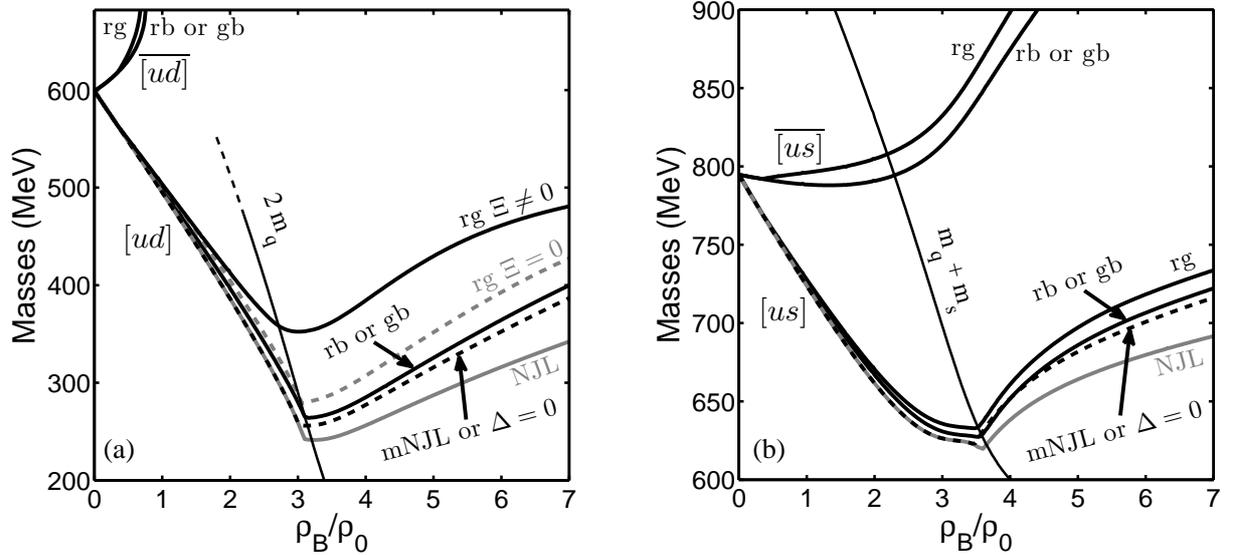

**FIG 14.** Masses of the scalar (anti)diquarks (a) $[ud]$, $\overline{[ud]}$, (b) $[us]$ and $\overline{[us]}$ according to $\rho_B$, at $T = 10$ MeV.

A clear distinction is visible in FIG 14(a) between the stable and resonant states: they are separated by the $2m_q$ curve and by an angular point. This affirmation does not concern the $rg\ \Xi \neq 0$ curve, because of its smooth evolution. More precisely, when $\Xi_{ud}$ is taken into account, $[ud]^{rg}$ becomes a resonance as soon as $\Delta_{ud} \neq 0$, i.e. when $\rho_B > 0.34\rho_0$ at $T = 10$ MeV. The FIG 15(a) confirms this affirmation and reveals that its width is strong at high densities. This behavior recalls the observations done for FIG 13, in which the influence of the 2SC phase on the meson widths is mentioned. Furthermore, Refs. [58,61,62] have reported the existence of a diquark resonance in this phase, with a non-negligible mass and width. Also, the NJL Ref. [51] explains that the light diquarks cease to exist as bound states when the chiral symmetry is restored, i.e. in the 2SC phase at low $T$.

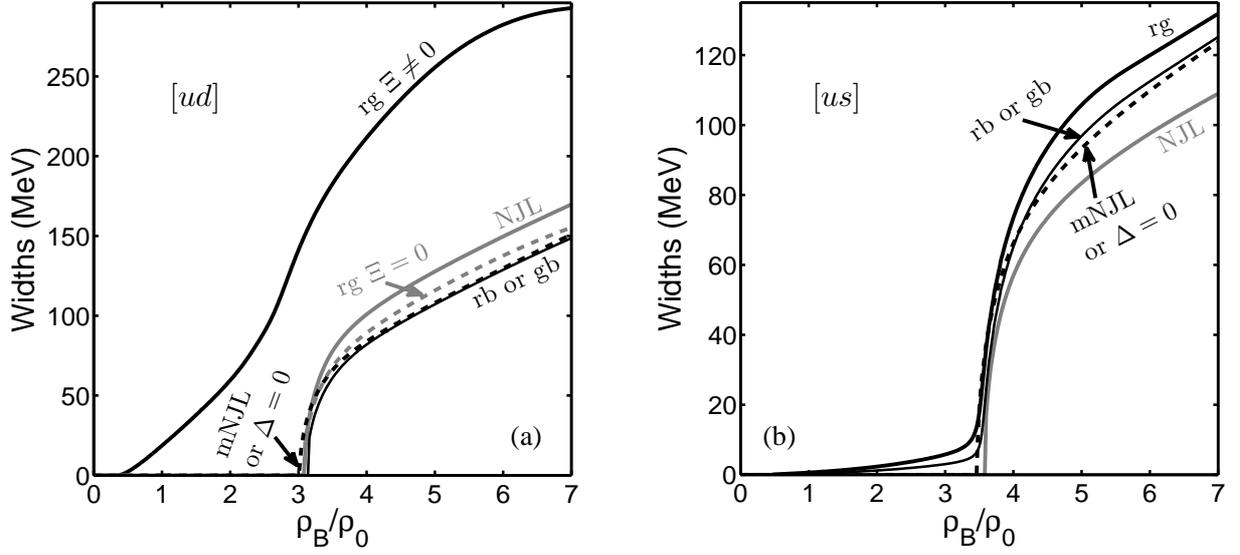

**FIG 15.** Widths of the scalar diquarks (a) $[ud]$ and (b) $[us]$ according to $\rho_B$, at $T = 10$ MeV.

Physically, the instability of $[ud]^{rg}$ in the 2SC phase can be understood as a manifestation of the formation of the diquark condensate. Indeed, the diquarks $[ud]^{rg}$ cease to be free and independent particles to form this single collective mode. Since $[ud]^{rg}$, $[ud]^{rb}$ and $[ud]^{gb}$ cannot mix in this modeling, the two last ones are not concerned by this phenomenon: they keep the behavior of particles. Only the indirect effect of the 2SC phase on the propagators of the red and green $q$ quarks is able to affect their masses. In the CFL phase, the diquarks $[ud]^{rg}$, $[us]^{rb}$ and $[ds]^{gb}$ could undergo the described phenomenon and become resonances in the same manner. Because of the inter-diquark coupling mentioned in Refs. [63,64], the other diquarks could be affected, but in a non-trivial way.

Moreover, the modeling described in Sec. III.C.1 also allows studying the antidiquarks. As expected, $[ud]$ and $\overline{[ud]}$ are only degenerate at null density. The masses of the antidiquarks are growing with the density, in agreement with Refs. [53,54]. A similar behavior is also found in Refs. [58,61,62], even if this increase is found in these papers according to $\mu_q$. As with the diquarks, a lifting of degeneracy between $\overline{[ud]}^{rg}$ and the two other antidiquarks $\overline{[ud]}^{rb}$ and $\overline{[ud]}^{gb}$ is visible in FIG 14(a). In addition, since the $\overline{[ud]}$ are made of two light antiquarks, they are strongly affected by the density. Therefore, they cannot be modeled after $\rho_0$.

Except for the effects of $\Xi_{ud}$, all the previous descriptions are qualitatively valid with $[us]$ and $\overline{[us]}$. Since they are formed of one strange quark/antiquark, their sensibility with respect to $\rho_B$ is lower, FIG 14(b). Consequently, $\overline{[us]}$ can be described for densities exceeding $3\rho_0$. Also, whatever the version of the model, the diquarks $[us]$ are stable at low densities, and become resonances when $\text{Re}(m_{[us]}) > m_q + m_s$, i.e. when $\rho_B > 3.5\rho_0$. However, because of the 2SC phase, the widths $\Gamma_{[us]}$ are non-zero (but weak) at lower $\rho_B$, FIG 15(b). This feature has been indentified with $K^-$ in FIG 6(d), but not with $[ud]^{rb}$ and $[ud]^{gb}$ in FIG 15(a).

The pseudoscalar diquarks $[ud]$ and $[us]$ are studied in FIG 16. They are always resonances, whatever the temperature or the baryonic density, as in Refs [53,54]. The previous comparisons between the various versions of the model are confirmed with these diquarks. Furthermore, as with the mesons, the restoration of the chiral symmetry at high densities leads to a degeneracy of the scalar and pseudoscalar $[ud]^{rg}$, and so on for the other diquarks.

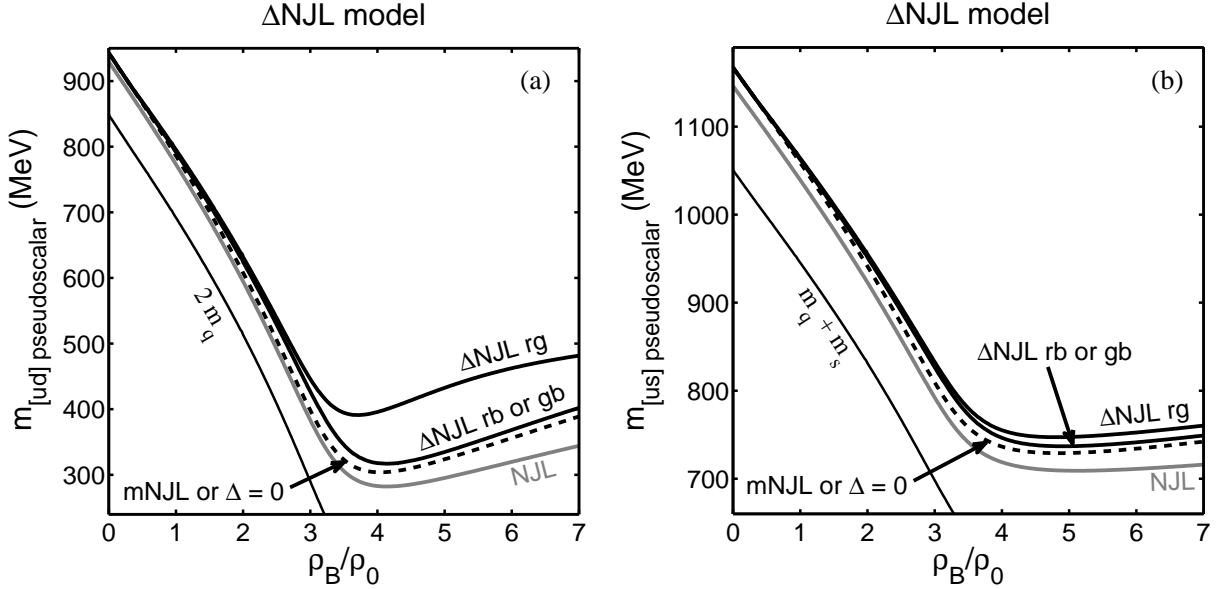

**FIG 16.** Masses of the pseudoscalar diquarks (a) $[ud]$ and (b) $[us]$ according to the baryonic density $\rho_B$, at $T = 10$ MeV.

In FIG 17 to FIG 19, the results obtained with the scalar diquarks are extended at finite temperatures. In the $\rho_B, T$ plane, the gray curve indicates the equality between the mass $m_{[ff']}$ of a diquark and the mass $m_f + m_{f'}$ of the quark pair that constitutes it. Except for $[ud]^{rg}$, this curve delimitates the stable and resonant regimes. In other words, it materializes the limit of the zone in which $\text{Im}(m_{[ff']}) \approx 0$. In the $\rho_B, T$ and $\mu_q, T$ planes, this transition is easily observable, since the curve is formed of angular points in the $m_{[ff']}$ surfaces. This confirms the results of Refs. [53,54], which present similar curves in the NJL description. Concerning $[ud]^{rg}$, the previous observations are only verified in the NQ phase. Indeed, this diquark is a resonance in the entire 2SC phase. This feature is in agreement with the interpretation of its behavior performed upstream at $T = 10$ MeV.

In the $\rho_B, T$ plane, the masses evolve continuously. In addition, at low $T$ and when $\mu_q < \mu_c$, the isodensity curves visible in the $\mu_q, T$ plane reveals that the baryonic density varies very slowly according to $\mu_q$. At low $T$, these two observations explain why the masses are rather constant according to $\mu_q$ when the chiral symmetry is broken. This remark concerns the quarks in FIG 4(a), the mesons in FIG 8(b) to FIG 11(b), and the diquarks in FIG 17(b) to FIG 19(b). Such a behavior is also visible in Refs [51,52]. These papers have performed an NJL description of the diquarks via loop functions, at finite temperatures and chemical potentials such as $\mu_q < \mu_c$.

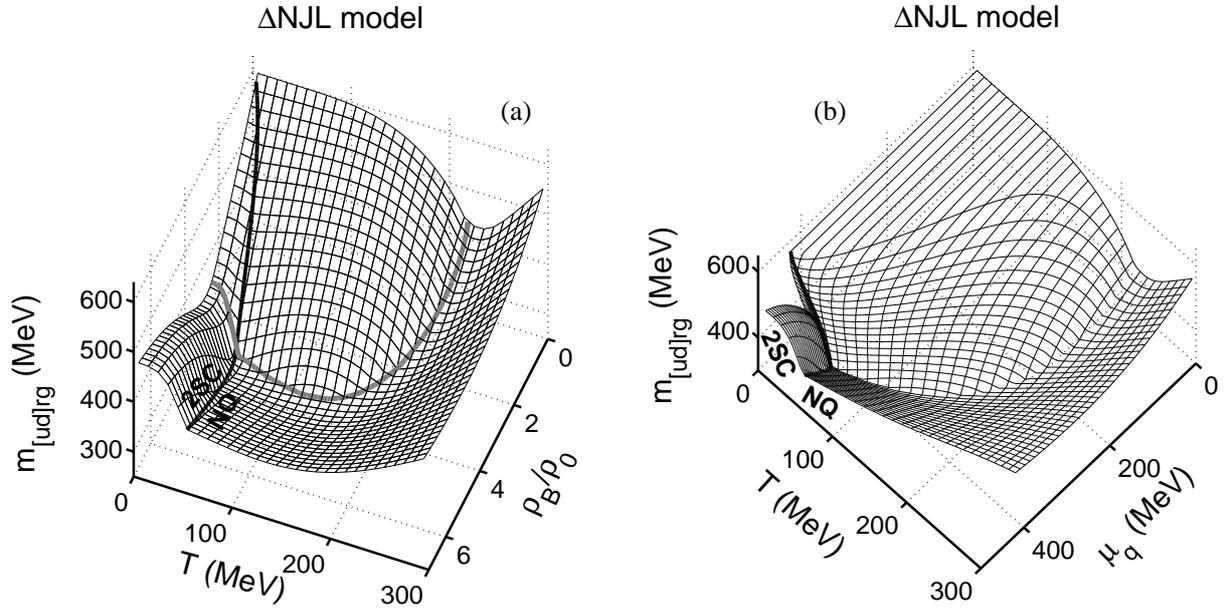

**FIG 17.** Mass of the scalar diquark $[ud]^{rg}$ in the (a) $\rho_B, T$ and (b) $\mu_q, T$ planes. The gray curve indicates when $m_{[ud]rg} = 2m_q$.

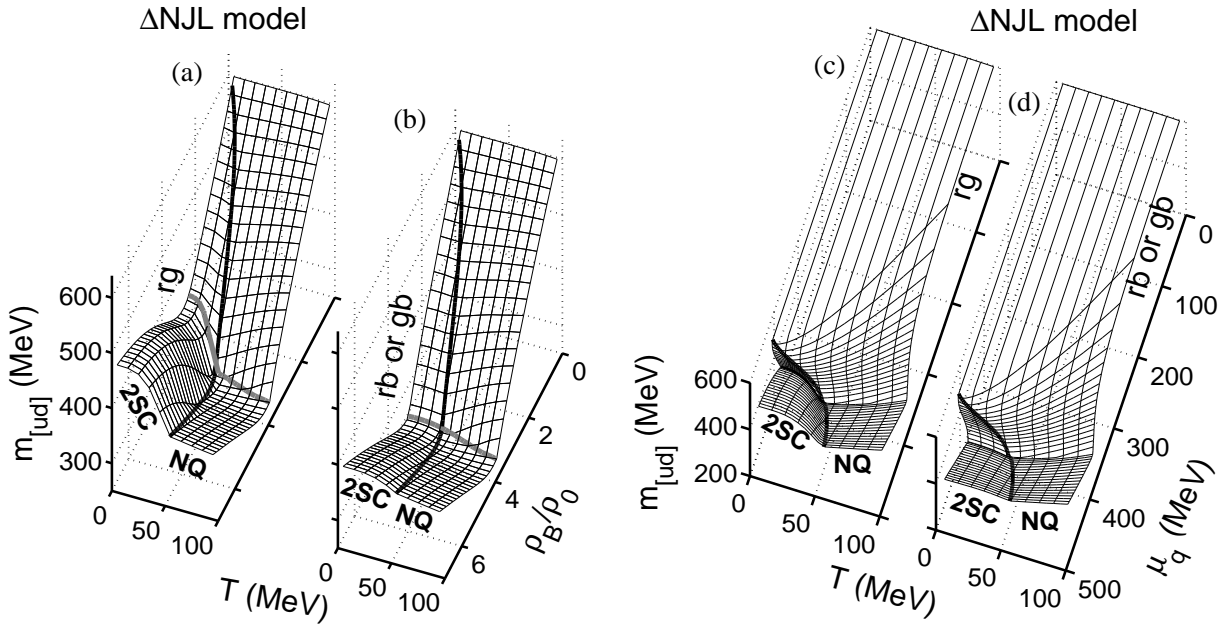

**FIG 18.** Mass of the scalar diquark $[ud]^{rg}$ in the (a) $\rho_B, T$ and (c) $\mu_q, T$ planes. Mass of $[ud]^{rb}$ or $[ud]^{gb}$ in the (b) $\rho_B, T$ and (d) $\mu_q, T$ planes.

In contrast, in Refs [58,61,62], the diquark masses significantly decrease according to the chemical potential until the chiral restoration. When $\mu_q \approx \mu_c$, they are close to zero. Massless or quasi-massless diquarks states are found in the chirally restored regime, in disagreement with the results found in this paper. The parametrization of the model cannot explain these differences, which necessarily come from the $\mu$ dependence of the diquark equations. However, concerning the heavy resonant state mentioned in these Refs., its mass increases with $\mu_q$ when $\mu_q > \mu_c$, in agreement with the lowest isotemperature curves visible in FIG 17(b).

The masses of $[ud]^{rg}$, $[ud]^{rb}$ and $[ud]^{gb}$ strictly coincide in the NQ phase. Consequently, FIG 18 focuses on a comparison in the 2SC phase of these diquarks, in the $\rho_B,T$ and $\mu_q,T$ planes. In agreement with FIG 14(a), the increase in the mass of $[ud]^{rb}$ and $[ud]^{gb}$ due to the 2SC phase is rather modest, since the $\Xi_{ud}$ terms are null for them. Nevertheless, this increase is strong enough to be visible in the graphs.

In FIG 19, the evolution of the $[us]^{rg}$ mass is in agreement with the NJL literature [53,54]. The 2SC phase only affects this composite particle via the $u$ quark propagator in the diquark loop function. Indeed, this $u$ quark is red or green, so necessarily affected by this phase, unlike the $s$ quark. This leads to a slight increase in the $[us]^{rg}$ mass. In the diquark $[us]^{rb}$ and $[us]^{gb}$, the $u$ quark is red/green or blue, i.e., respectively, affected or unaffected by the 2SC phase. These two possibilities are included in the modeling, Eqs. (63) and (64). They produce a weaker effect in comparison with $[us]^{rg}$. This explains the hierarchy observed in FIG 14(b). The masses of $[us]^{rb}$ and $[us]^{gb}$ in the $\rho_B,T$ and $\mu_q,T$ planes are easily deduced from these remarks and FIG 19.

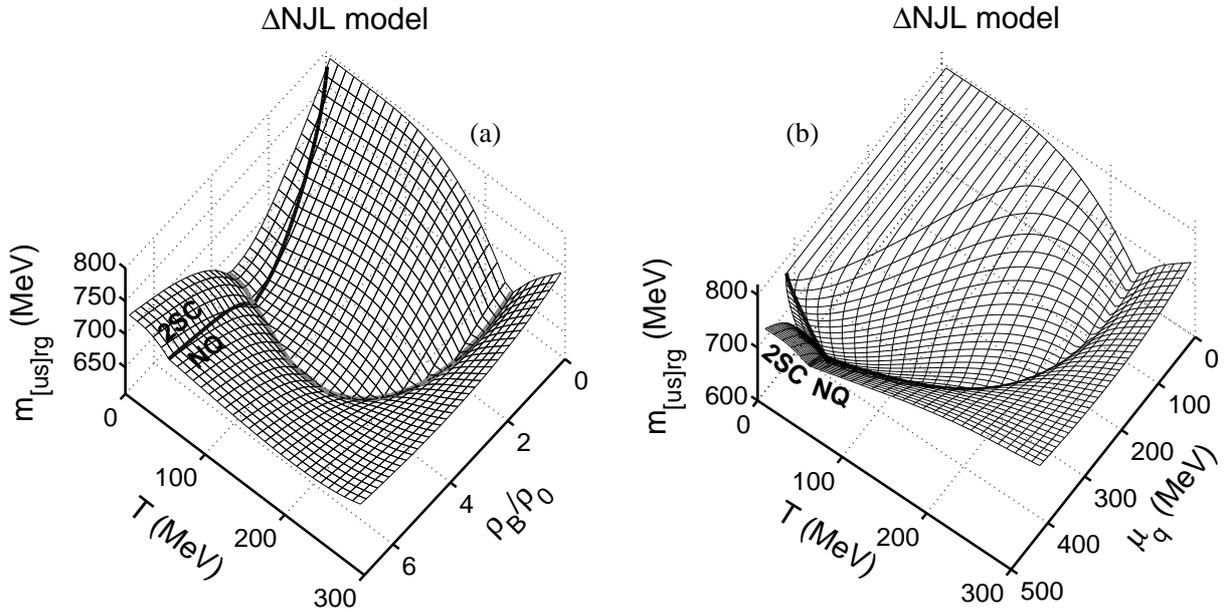

**FIG 19.** Mass of the scalar diquark $[us]^{rg}$ in the (a) $\rho_B,T$ and (b) $\mu_q,T$ planes. The gray curve indicates $m_{[us]rg} = m_q + m_s$.

*2. Inclusion of the Polyakov loop*

The only modification required to transform NJL-meson equations into PNJL ones is the replacement of the Fermi-Dirac distributions by $f_\Phi^\pm (E_f \mp \mu_f)$ [32], Eq. (54). This property is permitted by the color sum in Eq. (41) and by the relation $\mu_1^c - \mu_2^c = \mu_1 - \mu_2$. This subtraction appears in the term in front of $B_0$ in Eq. (51) and also during the evaluation of this function [57].

In the PNJL description of the diquarks, $\mu_1^c - \mu_2^c$ is replaced by $\mu_1^c + \mu_2^{c'}$ and there is no sum over the colors. Consequently, two approximations have been performed in Refs. [53,54] to treat the PNJL diquarks and mesons in the same way, i.e. with the replacement $f_{FD}(E_f \mp \mu_f) \to f_\Phi^\pm(E_f \mp \mu_f)$. In Eq. (69), the first consists in writing $\mu_1^c + \mu_2^{c'} \approx \mu_1 + \mu_2 \in \mathbb{R}$, including during the estimation of $B_0$. Indeed, the original version of $B_0$ only admits real arguments [57]. The second simplification consists in describing the diquarks $[ff']$ as an averaging over $[ff']^{rg}$, $[ff']^{rb}$ and $[ff']^{gb}$, with $ff' = ud$, $us$ or $ds$. Their degeneracy in the NQ phase found with NJL justifies this treatment. As far as I know, recent papers have not yet described PNJL diquarks without these simplifications. Therefore, it appears relevant to investigate their influence on the results. In the graphs presented hereafter, only the $(\Delta)\mu$PNJL models are employed, but all the observations stay qualitatively similar with $(\Delta)$PNJL. The only difference is the shifting of the values towards higher $T$.

The scalar diquarks are studied at null density in FIG 20. The color superconductivity cannot intervene in these conditions. The "$\mu$PNJL" curve has been obtained via the mentioned approximations, unlike the other curves found with $\Delta\mu$PNJL. In this model, the most striking result is the lifting of degeneracy between $[ud]^{rb}$ and the two other diquarks $[ud]$, and idem with the $[us]$. This feature is due to the definition of the $A_4$ matrix, i.e. $\beta A_4 = \phi_3 \lambda_3 + \phi_8 \lambda_8$, Sec. II.A.3. This leads to the following rewriting of Eq. (16)

$$\mu_f^r = \mu_f - iT(\phi_3 + \phi_8/\sqrt{3}), \quad \mu_f^g = \mu_f - iT(-\phi_3 + \phi_8/\sqrt{3}), \quad \mu_f^b = \mu_f - iT(-2\phi_8/\sqrt{3}). \quad (73)$$

Numerical calculations reveal that $\phi_3 \approx \phi_8/\sqrt{3}$. Consequently, $\text{Im}(\mu_f^g) \approx 0$ and $\text{Im}(\mu_f^b) \approx -\text{Im}(\mu_f^r)$. The minus sign in this second relation has no consequence on the diquark masses. So, the masses of $[ff']^{rg}$ and $[ff']^{gb}$ are identical, unlike those of $[ff']^{rb}$ that are slightly lower when $100 \text{ MeV} \leq T \leq 300 \text{ MeV}$. A different definition of the $A_4$ matrix leads to modifications of the results like, e.g., an exchange of the diquark curves. This phenomenon concerns colored composite particles, like the diquarks, and not the mesons or the baryons. Furthermore, when the diquarks are stable, the average of the three $[ff']^{cc'}$ masses strictly coincides with the $\mu$PNJL results. This feature is an argument in favor of the simplifications of Refs. [53,54]. In contrast, this property is no longer verified when the diquarks become resonances, because of the approximation due to the $B_0$ function.

All these observations stay valid at finite $\rho_B$, FIG 21. However, in these conditions, the imaginary part of the masses $m$ evolves differently. The $|\text{Im}(m)|$ curves of $[ff']^{rg}$ and $[ff']^{gb}$ exhibit significant non-zero values largely before reaching the resonant regime. This behavior is particularly visible at temperatures around 100–200 MeV and at moderate $\rho_B$. In contrast, this observation does not concern the $[ff']^{rb}$. Indeed, the property $\text{Im}(\mu_f^b) \approx -\text{Im}(\mu_f^r)$ allows the relation $\mu_1^r + \mu_2^b \approx \mu_1 + \mu_2 \in \mathbb{R}$ to be verified, unlike $\mu_1^r + \mu_2^g$ or $\mu_1^g + \mu_2^b$ that are complex. So, the equations (and their solutions) associated with the diquarks $[ff']^{rb}$ stay real when these particles are stable.

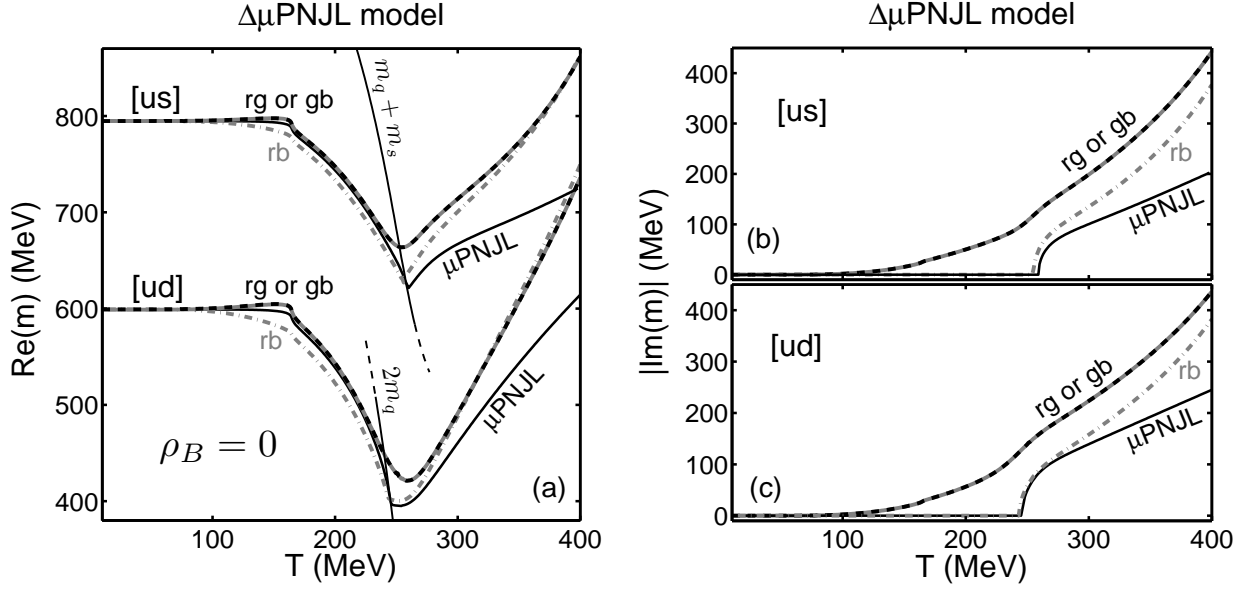

**FIG 20.** (a) Real part of the masses of the scalar diquarks $[ud]$ and $[us]$, according to the temperature $T$. Absolute value of the corresponding imaginary part of the (b) $[us]$ and (c) $[ud]$ diquark masses. The baryonic density is fixed to zero.

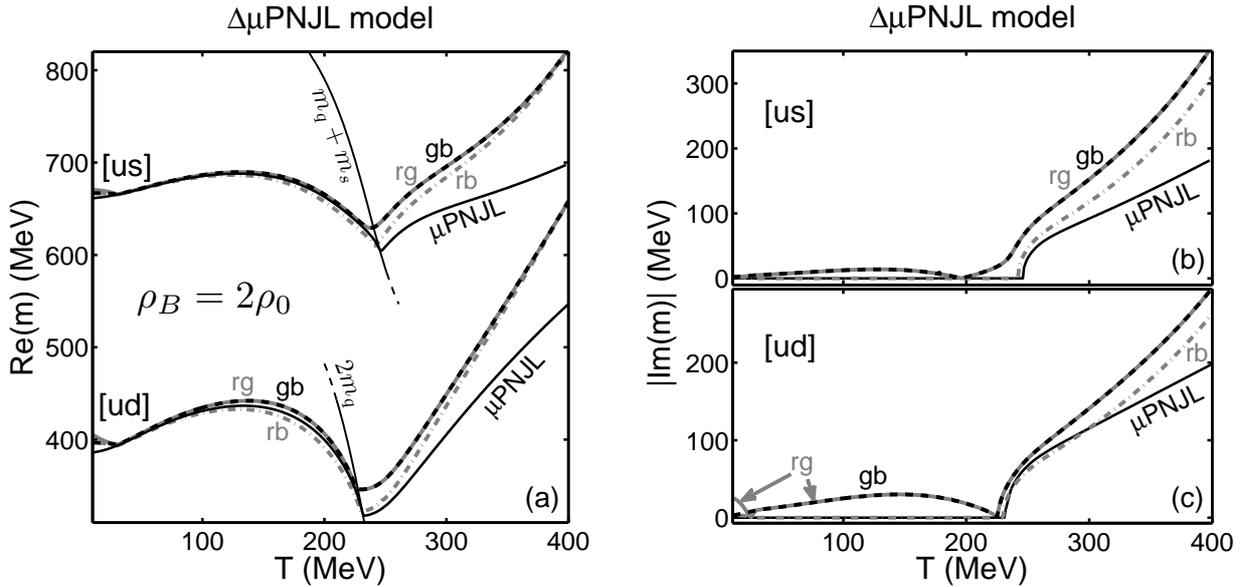

**FIG 21.** (a) Real part of the $[ud]$ and $[us]$ diquark masses, according to the temperature $T$ and at $\rho_B = 2\rho_0$. Imaginary part of the (b) $[us]$ and (c) $[ud]$ diquark masses.

Moreover, $[ff']^{rg}$ and $[ff']^{gb}$ are no longer degenerate at the coldest temperatures of FIG 21, because of the 2SC phase that intervenes at this density when $T < 30$ MeV. In FIG 22, this situation is reversed, since this phase is present at $T = 10$ MeV when $\rho_B > 0.3\rho_0$. At this temperature, the lifting of degeneracy described in the two previous figures is strongly reduced. In contrast, the effect of the color superconductivity is easily visible: it breaks the symmetry between $[ff']^{rg}$ and the two other diquarks $[ff']^{rb}$ and $[ff']^{gb}$. The color superconductivity involves low temperatures. Therefore, these observations are generalizable: the definition of the $A_4$ matrix is without notable consequence on the diquark masses in the 2SC phase, at least for the densities handled in this paper. In addition, FIG 22 has strong similarities with FIG 14 and FIG 15. For example, the values reached by the width of $[ud]^{rg}$ in FIG 15 are found again, via a factor 2 because of Eq. (60). Nevertheless, when the diquarks are stable, the behavior of the $|\text{Im}(m)|$ curves constitutes a difference.

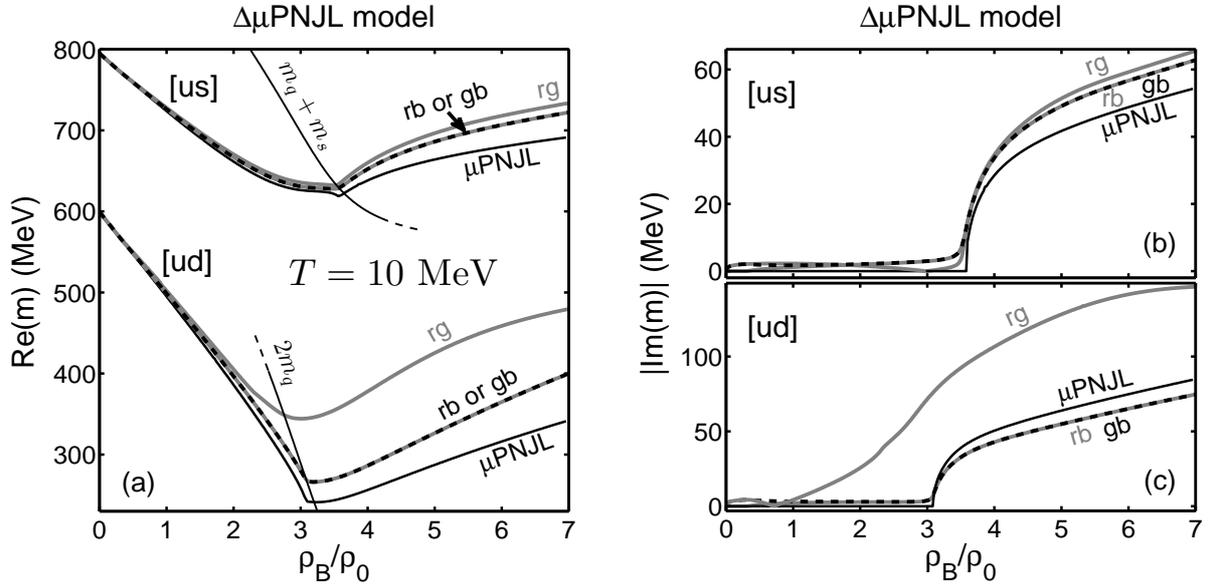

**FIG 22.** (a) Real part of the $[ud]$ and $[us]$ diquark masses, according to the baryonic density $\rho_B$, at $T = 10$ MeV. Imaginary part of the (b) $[us]$ and (c) $[ud]$ diquark masses.

## V. CONCLUSIONS

In this paper, I have investigated the pseudoscalar/scalar mesons and diquarks at finite temperatures $T$, baryonic densities $\rho_B$ and chemical potentials $\mu_q$. This includes the description of their behavior in a color superconducting regime. To reach this objective, various versions of the NJL model (notably including a Polyakov loop) have been considered to study these composite particles in a three-flavor and three-color approach. This modeling has been done via an adaptation at finite temperatures of the formalism described in Refs. [58-62]. During the establishment of the equations, it has been observed some common points with pure NJL [27,50] or PNJL [32,34,53,54] descriptions. However, the inclusion of the color superconductivity has imposed to build new algorithms and to go beyond certain approximations usually performed in the standard NJL approach [57].

In a first time, the influence on the quarks of the color superconductivity and meson condensation has been indicated. The meson condensation has a strong impact on the quark masses. This lets foresee that this phenomenon certainly has a non-negligible effect on the composite particles. This observation is probably extensible at finite strangeness. Nevertheless, with the isospin symmetry and at null strangeness, the color symmetric phase (normal quark phase) and the 2SC phase are the only relevant ones. Indeed, the pion condensation requires a strong asymmetry between the $u$ and $d$ quarks and the other phases need $\mu_s \neq 0$. Therefore, a possible extension of this work may concern a modeling performed beyond the isospin symmetry and at finite strangeness. Since the method described in Ref. [45] is restrained to the framework of the isospin symmetry, the relations gathered in Appendix A should be useful in this task. Such a development could allow studying the influence of the pseudoscalar meson condensation on the mesons [85] and diquarks. In addition, finite strangeness could allow taking into account the CFL and kaon condensation phases [42].

Then, one has focused on the description of the mesons and diquarks. It has been concluded that the approximations performed in previous works [53,54,57] to describe these particles with (P)NJL are without great consequences on the estimated masses. Nevertheless, the widths are sometimes underestimated because of certain of these simplifications. Also, the equations established in this paper have permitted to find results with a good agreement with the (P)NJL literature, in the normal quark and 2SC phases. It has been verified that the chiral restoration at high densities leads to a degeneracy of $\pi$ and $f_0$, and so on for the other mesons and for the diquarks, even in the color superconducting regime. However, this feature may be altered by the diquark/meson coupling, which is neglected in this document. So, a possible development may include this coupling [58,65,66].

In addition, it has been observed the strong influence of the 2SC phase on the pions, $\eta$ and the diquark $[ud]$ whose color is $rg$. A striking result is the fact that this diquark is a resonance in all this color superconducting phase. This behavior has been explained via the formation of the diquark condensate, which prevents $[ud]^{rg}$ to act as a free particle. In contrast, the composite particles made of one strange quark/antiquark are weakly affected by the 2SC phase. However, this phase leads certain mesons and diquarks to have a weak but non-zero width in their stability regime. The modifications leaded by the Polyakov loop do not contradict all these observations. Anyway, the inclusion of the 2SC phase in the modeling appears essential to study mesons and diquarks at low temperatures and at finite baryonic densities. This remark includes their description near the first order chiral phase transition.

Finally, the presented work opens the door to two possible investigations. Firstly, the study of the diquarks is motivated by the baryon modeling via a quark-diquark picture. Consequently, a direct continuation of this paper may consist in a description of the baryons in a color superconducting regime [86]. Secondly, this concerns the study of reactions involving mesons, diquarks and baryons, via the calculations of the corresponding cross-sections. The relevance of reactions like $q+\bar{q} \to M+M$, $q+q \to M+D$, $D+D \to q+B$, $q+q \to \bar{q}+B$ and $q+D \to M+B$ on the mesonization or baryonization processes has been underlined [55,56] ($q$, $D$, $M$ and $B$ stand for, respectively, quark, diquark, meson and baryon). It should be instructive to investigate the influence of the 2SC phase on such reactions. If its effect is strong enough, it may constitute a proof of the formation of the color superconductivity in experiments…

## APPENDIX A: PROPAGATORS OF THE QUARKS AND ANTIQUARKS

In this Appendix, the subscripts refer to the flavor, i.e. *u, d* or *s*. The *f* indicates an unspecified flavor. The superscript sign is a plus for a quark or a minus for an antiquark. The color is also mentioned in superscript. It may be $r, g, b$ or $c$ in the case of an unspecified color. The various expressions of the quark/antiquark propagators are listed hereafter, for each phase. This description does not respect the isospin symmetry.

### 1. The normal quark (NQ) phase

In this color symmetric phase, the $S^{\pm}$ matrices are block diagonal. The propagators are then written with the well-known expression

$$S_f^{c\pm} = \frac{1}{\displaystyle{\not{p} \pm \gamma_0 \mu_f^c - m_f}}.\tag{A1}$$

The $\Xi^{\pm}$ matrices that translate the quark-quark pairing are null,

$$\Xi_{ff'}^{cc'\pm} = 0.\tag{A2}$$

## 2. The pion condensation phase

The $S^{\pm}$ matrices are now written in the form

$$S^{\pm} = \begin{bmatrix} S_u^{\pm} & \Theta_{ud}^{\pm} & 0 \\ \Theta_{du}^{\pm} & S_d^{\pm} & 0 \\ 0 & 0 & S_s^{\pm} \end{bmatrix}. \tag{A3}$$

The propagators for the $u$ quarks/antiquarks are

$$S_u^{c\pm} = \frac{1}{\slashed{p} \pm \gamma_0 \mu_u^c - m_u + \pi_d^{c\pm}\left(\slashed{p} \pm \gamma_0 \mu_d^c - m_d\right)} \quad \text{with} \quad \pi_d^{c\pm} = \frac{-(\Delta_\pi)^2}{\left(i\omega_n \pm \mu_d^c\right)^2 - E_d^2}. \tag{A4}$$

The expressions for the $d$ quarks/antiquarks are found with the property

$$S_d^{c\pm} = S_u^{c\pm}\bigg|_{u \leftrightarrow d}. \tag{A5}$$

Since the pion condensate does not affect the strange quarks/antiquarks,

$$S_s^{c\pm} = \frac{1}{\slashed{p} \pm \gamma_0 \mu_s^c - m_s}. \tag{A6}$$

Concerning the $\Theta^{\pm}$ terms that translate the quark-antiquark coupling,

$$\Theta_{ud}^{cc\pm} = \pm S_u^{c\pm}\left(\Delta_\pi \gamma_5\right) \frac{1}{\slashed{p} \pm \gamma_0 \mu_d^c - m_d} \quad \text{and} \quad \Theta_{du}^{cc\pm} = -\Theta_{ud}^{cc\pm}\bigg|_{u \leftrightarrow d}. \tag{A7}$$

As in the normal quark phase, the quark-quark pairing is null, leading to $\Xi_{ff'}^{cc'\pm} = 0$.

## 3. The 2-flavor color-superconducting (2SC) phase

The $S^{\pm}$ matrices are block diagonal in this phase, since $\Theta_{ff'}^{cc\pm} = 0$. The propagators of the red $u$ quarks/antiquarks are

$$S_u^{r\pm} = \frac{1}{\slashed{p} \pm \gamma_0 \mu_u^r - m_u + \delta_d^{g\pm}\left(\slashed{p} \mp \gamma_0 \mu_d^g - m_d\right)} \quad \text{with} \quad \delta_d^{g\pm} = \frac{-|\Delta_{ud}|^2}{\left(i\omega_n \mp \mu_d^g\right)^2 - E_d^2}. \tag{A8}$$

The propagators of the green $u$ quarks/antiquarks satisfy the property

$$S_u^{g\pm} = S_u^{r\pm}\bigg|_{r \leftrightarrow g}. \tag{A9}$$

In the 2SC phase, the red $u$ quarks interact with the green $d$, and the green $u$ with the red $d$. In contrast, the blue $u$ and $d$ are uncoupled, as well as the strange quarks. Furthermore, the expressions for the $d$ quarks/antiquarks are obtained via the relation

$$S_d^{c\pm} = S_u^{c\pm}\bigg|_{u \leftrightarrow d}. \tag{A10}$$

Concerning the non-null coupling terms, one writes

$$\Xi_{ud}^{rg-} = S_u^{r+}\left(\Delta_{ud}\gamma_5\right)\frac{1}{\slashed{p} - \gamma_0 \mu_d^g - m_d}, \quad \Xi_{ud}^{rg+} = S_u^{r-}\left(-\Delta_{ud}^*\gamma_5\right)\frac{1}{\slashed{p} + \gamma_0 \mu_d^g - m_d}, \tag{A11}$$

and one has

$$\Xi_{ud}^{gr\pm} = -\Xi_{ud}^{rg\pm}\bigg|_{r \leftrightarrow g} \quad \text{and} \quad \Xi_{du}^{rg\pm} = -\Xi_{ud}^{rg\pm}\bigg|_{d \leftrightarrow u}. \tag{A12}$$

## APPENDIX B: ENERGY PROJECTORS

### 1. Definition and properties

The first vocation is this Appendix is to recall the formulas of the energy projectors introduced in Ref. [45], and to extend these relations. The energy projectors are defined as

$$\overline{\Lambda}_f^{\pm}(\vec{p}) = \frac{1}{2}\left[1 \pm \frac{\gamma_0(\vec{\gamma}\cdot\vec{p} - m_f)}{E_f}\right] \text{ and } \Lambda_f^{\pm}(\vec{p}) = \frac{1}{2}\left[1 \pm \frac{\gamma_0(\vec{\gamma}\cdot\vec{p} + m_f)}{E_f}\right], \quad (B1)$$

where $E_f = \sqrt{\vec{p}^2 + m_f^2}$. They verify the following properties

$$\begin{cases} \overline{\Lambda}_f^{\pm}(\vec{p})\overline{\Lambda}_f^{\pm}(\vec{p}) = \overline{\Lambda}_f^{\pm}(\vec{p}) \\ \overline{\Lambda}_f^{\pm}(\vec{p})\overline{\Lambda}_f^{\mp}(\vec{p}) = 0 \quad \text{and so on for } \Lambda_f^{\pm}(\vec{p}). \\ \overline{\Lambda}_f^{+}(\vec{p}) + \overline{\Lambda}_f^{-}(\vec{p}) = 1 \end{cases} \quad (B2)$$

In addition, with the matrices $\gamma_0$ and $\gamma_5$,

$$\begin{cases} \gamma_0 \overline{\Lambda}_f^{\mp}(\vec{p}) \gamma_0 = \Lambda_f^{\pm}(\vec{p}) \\ \gamma_5 \overline{\Lambda}_f^{\pm}(\vec{p}) \gamma_5 = \Lambda_f^{\pm}(\vec{p}) \end{cases} \text{ and } \begin{cases} \gamma_0 \Lambda_f^{\pm}(\vec{p}) \gamma_0 = \overline{\Lambda}_f^{\mp}(\vec{p}) \\ \gamma_5 \Lambda_f^{\pm}(\vec{p}) \gamma_5 = \overline{\Lambda}_f^{\pm}(\vec{p}) \end{cases}. \quad (B3)$$

Concerning the trace calculations, one has

$$Tr\left[\overline{\Lambda}_f^{\pm}(\vec{p})\right] = Tr\left[\Lambda_f^{\pm}(\vec{p})\right] = 2. \quad (B4)$$

Moreover, one gets

$$Tr\left[\Lambda_f^{\pm}(\vec{p})\overline{\Lambda}_{f'}^{\pm}(\vec{p}')\right] = 1 + \frac{\vec{p}\cdot\vec{p}' - m_f m_{f'}}{E_f E_{f'}} \text{ and } Tr\left[\Lambda_f^{\pm}(\vec{p})\overline{\Lambda}_{f'}^{\mp}(\vec{p}')\right] = 1 - \frac{\vec{p}\cdot\vec{p}' - m_f m_{f'}}{E_f E_{f'}}, \quad (B5)$$

which give, respectively, $2\vec{p}^2/E_f^2$ and $2m_f^2/E_f^2$ when $f = f'$ and $\vec{p} = \vec{p}'$ [59]. In addition,

$$Tr\left[\overline{\Lambda}_f^{\pm}(\vec{p})\overline{\Lambda}_{f'}^{\pm}(\vec{p}')\right] = Tr\left[\Lambda_f^{\pm}(\vec{p})\Lambda_{f'}^{\pm}(\vec{p}')\right] = 1 + \frac{\vec{p}\cdot\vec{p}' + m_f m_{f'}}{E_f E_{f'}} \quad (B6)$$

and

$$Tr\left[\overline{\Lambda}_f^{\pm}(\vec{p})\overline{\Lambda}_{f'}^{\mp}(\vec{p}')\right] = Tr\left[\Lambda_f^{\pm}(\vec{p})\Lambda_{f'}^{\mp}(\vec{p}')\right] = 1 - \frac{\vec{p}\cdot\vec{p}' + m_f m_{f'}}{E_f E_{f'}}. \quad (B7)$$

### 2. Propagators of the quarks and antiquarks

The energy projectors allow rewriting the propagators listed in the previous Appendix. However, concerning the 2SC or pion condensation phases, this method is only applicable with the isospin symmetry. In the normal quark phase, the propagators of the quarks/antiquarks take the form

$$S_f^{c\pm} = \frac{\gamma_0 \overline{\Lambda}_f^{+}(\vec{p})}{i\omega_n + E_f \pm \mu_f^c} + \frac{\gamma_0 \overline{\Lambda}_f^{-}(\vec{p})}{i\omega_n - E_f \pm \mu_f^c}. \quad (B8)$$

In the pion condensation phase, they become for the light quarks

$$S_q^{c\pm} = \frac{\left(i\omega_n \pm \mu_q^c - E_q\right)\gamma_0 \overline{\Lambda}_q^{+}(\vec{p}) + \left(i\omega_n \pm \mu_q^c + E_q\right)\gamma_0 \overline{\Lambda}_q^{-}(\vec{p})}{\left(i\omega_n \pm \mu_q^c\right)^2 - E_q^2 - \left(\Delta_\pi\right)^2}. \quad (B9)$$

Equation (A7) is rewritten as

$$\Theta_{ud}^{cc\pm} = \mp \Delta_\pi \frac{\gamma_5 \overline{\Lambda}_q^{+}(\vec{p}) + \gamma_5 \overline{\Lambda}_q^{-}(\vec{p})}{\left(i\omega_n \pm \mu_q^c\right)^2 - E_q^2 - \left(\Delta_\pi\right)^2}. \quad (B10)$$

Moreover, in the 2SC phase, Eq. (A8) is now

$$S_q^{r\pm} = \frac{\left(i\omega_n - E_q \mp \mu_q^g\right)\gamma_0 \bar{\Lambda}_q^+(\vec{p})}{\left(i\omega_n + E_q \pm \mu_q^r\right)\left(i\omega_n - E_q \mp \mu_q^g\right) - |\Delta_{ud}|^2} + \frac{\left(i\omega_n + E_q \mp \mu_q^g\right)\gamma_0 \bar{\Lambda}_q^-(\vec{p})}{\left(i\omega_n - E_q \pm \mu_q^r\right)\left(i\omega_n + E_q \mp \mu_q^g\right) - |\Delta_{ud}|^2}.$$ (B11)

In the same way, Eq. (A11) takes the form

$$\Xi_{ud}^{rg-} = \frac{-\Delta_{ud}\gamma_5 \bar{\Lambda}_q^+(\vec{p})}{\left(i\omega_n - E_q + \mu_q^r\right)\left(i\omega_n + E_q - \mu_q^g\right) - |\Delta_{ud}|^2} + \frac{-\Delta_{ud}\gamma_5 \bar{\Lambda}_q^-(\vec{p})}{\left(i\omega_n + E_q + \mu_q^r\right)\left(i\omega_n - E_q - \mu_q^g\right) - |\Delta_{ud}|^2}$$ (B12)

and

$$\Xi_{ud}^{rg+} = \frac{\Delta_{ud}^*\gamma_5 \bar{\Lambda}_q^+(\vec{p})}{\left(i\omega_n - E_q - \mu_q^r\right)\left(i\omega_n + E_q + \mu_q^g\right) - |\Delta_{ud}|^2} + \frac{\Delta_{ud}^*\gamma_5 \bar{\Lambda}_q^-(\vec{p})}{\left(i\omega_n + E_q - \mu_q^r\right)\left(i\omega_n - E_q + \mu_q^g\right) - |\Delta_{ud}|^2}.$$ (B13)

## APPENDIX C: DETAIL ON $\mathcal{S}_{Mixed}$

In Eq. (36), $\mathcal{S}_{Mixed}$ allows observing a coupling between certain pseudoscalar mesons and pseudoscalar diquarks on the one hand, and between certain scalar mesons and scalar diquarks on the other hand. Concerning the first kind of coupling, it appears terms gathering $\underline{\Delta}_{\eta 0}, \underline{\Delta}_{\pi^0}, \underline{\Delta}_{\eta 8}$ with $\underline{\Delta}_{ud(PS)}^{2\,*}$ and $\underline{\Delta}_{ud(PS)}^{2}$, $\underline{\Delta}_{K^+}$ with $\underline{\Delta}_{ds(PS)}^{2\,*}$, $\underline{\Delta}_{K^-}$ with $\underline{\Delta}_{ds(PS)}^{2}$, $\underline{\Delta}_{K^0}$ with $\underline{\Delta}_{us(PS)}^{2\,*}$ and $\underline{\Delta}_{\bar{K}^0}$ with $\underline{\Delta}_{us(PS)}^{2}$, where PS stands for pseudoscalar. Also, $\underline{\Delta}_{\eta 0} = \underline{\Delta}_0$ and $\underline{\Delta}_{\eta 8} = \underline{\Delta}_8$, Eqs. (32) and (33). In the same way, the second kind involves $\underline{\Delta}_{f_0 0}, \underline{\Delta}_{a_0^0}, \underline{\Delta}_{f_0 8}$ with $\underline{\Delta}_{ud}^{2\,*}$ and $\underline{\Delta}_{ud}^{2}$, $\underline{\Delta}_{K_0^{*+}}$ with $\underline{\Delta}_{ds}^{2\,*}$, $\underline{\Delta}_{K_0^{*-}}$ with $\underline{\Delta}_{ds}^{2}$, $\underline{\Delta}_{K_0^{*0}}$ with $\underline{\Delta}_{us}^{2\,*}$ and $\underline{\Delta}_{\bar{K}_0^{*0}}$ with $\underline{\Delta}_{us}^{2}$.

However, $\mathcal{S}_{Mixed}$ is greatly simplified with the isospin symmetry. In this case, the coupling between the pseudoscalar mesons and diquarks is expressed by the terms

$$\begin{aligned}
\frac{i}{2}Tr\Bigg[ & S_q^+ \left(\sqrt{\frac{2}{3}}\underline{\Delta}_{\eta 0} + \frac{1}{\sqrt{3}}\underline{\Delta}_{\eta 8}\right) i\gamma_5 \, \Xi_{ud}^- \, \underline{\Delta}_{ud(PS)}^{2\,*}\lambda_2 + \Xi_{ud}^- \, \underline{\Delta}_{ud(PS)}^{2\,*}\lambda_2 \, S_q^+ \left(\sqrt{\frac{2}{3}}\underline{\Delta}_{\eta 0} + \frac{1}{\sqrt{3}}\underline{\Delta}_{\eta 8}\right) i\gamma_5 \\
& - S_q^- \left(\sqrt{\frac{2}{3}}\underline{\Delta}_{\eta 0} + \frac{1}{\sqrt{3}}\underline{\Delta}_{\eta 8}\right) i\gamma_5 \, \Xi_{ud}^+ \, \underline{\Delta}_{ud(PS)}^{2}\lambda_2 - \Xi_{ud}^+ \, \underline{\Delta}_{ud(PS)}^{2}\lambda_2 \, S_q^- \left(\sqrt{\frac{2}{3}}\underline{\Delta}_{\eta 0} + \frac{1}{\sqrt{3}}\underline{\Delta}_{\eta 8}\right) i\gamma_5 \\
& + S_q^- \underline{\Delta}_{ud(PS)}^{2\,*}\lambda_2 \, \Xi_{ud}^- \left(\sqrt{\frac{2}{3}}\underline{\Delta}_{\eta 0} + \frac{1}{\sqrt{3}}\underline{\Delta}_{\eta 8}\right) i\gamma_5 + \Xi_{ud}^- \left(\sqrt{\frac{2}{3}}\underline{\Delta}_{\eta 0} + \frac{1}{\sqrt{3}}\underline{\Delta}_{\eta 8}\right) i\gamma_5 \, S_q^- \underline{\Delta}_{ud(PS)}^{2\,*}\lambda_2 \\
& - S_q^+ \underline{\Delta}_{ud(PS)}^{2}\lambda_2 \, \Xi_{ud}^+ \left(\sqrt{\frac{2}{3}}\underline{\Delta}_{\eta 0} + \frac{1}{\sqrt{3}}\underline{\Delta}_{\eta 8}\right) i\gamma_5 - \Xi_{ud}^+ \left(\sqrt{\frac{2}{3}}\underline{\Delta}_{\eta 0} + \frac{1}{\sqrt{3}}\underline{\Delta}_{\eta 8}\right) i\gamma_5 \, S_q^+ \underline{\Delta}_{ud(PS)}^{2}\lambda_2 \Bigg]
\end{aligned}$$ (C1)

and so on for the coupling between scalar mesons and diquarks. Therefore, $\underline{\Delta}_{\eta 0}, \underline{\Delta}_{\eta 8}$ with $\underline{\Delta}_{ud(PS)}^{2\,*}, \underline{\Delta}_{ud(PS)}^{2}$ translates a coupling between the $\eta, \eta'$ mesons and the diquark $[ud]_{PS}^{rg}$, and $\underline{\Delta}_{f_0 0}, \underline{\Delta}_{f_0 8}$ with $\underline{\Delta}_{ud}^{2\,*}, \underline{\Delta}_{ud}^{2}$ a coupling between $f_0, f_0'$ and the scalar diquark $[ud]^{rg}$.

## APPENDIX D: THE PSEUDOSCALAR MESONS $\pi^0, \eta, \eta'$

As in Refs. [53,54], I introduce the $\Pi$ matrix defined as

$$\Pi = \begin{bmatrix} \Pi_{00} & \Pi_{03} & \Pi_{08} \\ \Pi_{30} & \Pi_{33} & \Pi_{38} \\ \Pi_{80} & \Pi_{83} & \Pi_{88} \end{bmatrix} \tag{D1}$$

with

$$\Pi_{00} = \frac{-i}{4}Tr\left[\frac{2}{3}\left(2S_{u\bar{u}} + 2S_{d\bar{d}} + 2S_{s\bar{s}} + \Xi_{ud} + \Xi_{du}\right)\right], \quad \Pi_{33} = \frac{-i}{4}Tr\left[2S_{u\bar{u}} + 2S_{d\bar{d}} - \Xi_{ud} - \Xi_{du}\right],$$

$$\Pi_{88} = \frac{-i}{4}Tr\left[\frac{1}{3}\left(2S_{u\bar{u}} + 2S_{d\bar{d}} + 8S_{s\bar{s}} + \Xi_{ud} + \Xi_{du}\right)\right],$$

$$\Pi_{03} = \frac{-i}{4}Tr\left[\sqrt{\frac{2}{3}}\left(2S_{u\bar{u}} - 2S_{d\bar{d}} + \Xi_{ud} - \Xi_{du}\right)\right], \quad \Pi_{30} = \frac{-i}{4}Tr\left[\sqrt{\frac{2}{3}}\left(2S_{u\bar{u}} - 2S_{d\bar{d}} - \Xi_{ud} + \Xi_{du}\right)\right], \tag{D2}$$

$$\Pi_{08} = \Pi_{80} = \frac{-i}{4}Tr\left[\frac{\sqrt{2}}{3}\left(2S_{u\bar{u}} + 2S_{d\bar{d}} - 4S_{s\bar{s}} + \Xi_{ud} + \Xi_{du}\right)\right],$$

$$\Pi_{38} = \frac{-i}{4}Tr\left[\frac{1}{\sqrt{3}}\left(2S_{u\bar{u}} - 2S_{d\bar{d}} - \Xi_{ud} + \Xi_{du}\right)\right], \quad \Pi_{83} = \frac{-i}{4}Tr\left[\frac{1}{\sqrt{3}}\left(2S_{u\bar{u}} - 2S_{d\bar{d}} + \Xi_{ud} - \Xi_{du}\right)\right],$$

with the shorthand notations

$$\Xi_{ud} = \Xi_{ud}^{rg-} i\gamma_5 \Xi_{du}^{gr+} i\gamma_5 + \Xi_{ud}^{gr-} i\gamma_5 \Xi_{du}^{rg+} i\gamma_5 + \Xi_{ud}^{rg+} i\gamma_5 \Xi_{du}^{gr-} i\gamma_5 + \Xi_{ud}^{gr+} i\gamma_5 \Xi_{du}^{rg-} i\gamma_5 \tag{D3}$$

and $\Xi_{du} = \Xi_{ud}\big|_{u \leftrightarrow d}$.

When the isospin symmetry is employed, $\Xi_{ud} = \Xi_{du} = -2\Xi_q$, where $\Xi_q$ is defined Eq. (46). This leads to the simplifications

$$\Pi_{88} = \frac{-i}{2}Tr\left[\frac{1}{3}\left(2S_{q\bar{q}} + 4S_{s\bar{s}} - 2\Xi_q\right)\right], \quad \Pi_{08} = \Pi_{80} = \frac{-i}{2}Tr\left[\frac{\sqrt{2}}{3}\left(2S_{q\bar{q}} - 2S_{s\bar{s}} - 2\Xi_q\right)\right], \tag{D4}$$

$\Pi_{38} = 0$, $\Pi_{83} = 0$.

The $\Pi$ matrix becomes

$$\Pi = \frac{-i}{2}\begin{bmatrix} \frac{2}{3}Tr\left(2S_{q\bar{q}} + S_{s\bar{s}} - 2\Xi_q\right) & 0 & \frac{2\sqrt{2}}{3}Tr\left(S_{q\bar{q}} - S_{s\bar{s}} - \Xi_q\right) \\ 0 & 2Tr\left(S_{q\bar{q}} + \Xi_q\right) & 0 \\ \frac{2\sqrt{2}}{3}Tr\left(S_{q\bar{q}} - S_{s\bar{s}} - \Xi_q\right) & 0 & \frac{2}{3}Tr\left(S_{q\bar{q}} + 2S_{s\bar{s}} - \Xi_q\right) \end{bmatrix}. \tag{D5}$$

Consequently, the pion $\pi^0$ (channel 3) is decoupled from $\eta, \eta'$ (channels 0 and 8) thanks to the isospin symmetry. Furthermore, $\pi^0$ corresponds to the $2Tr\left(S_{q\bar{q}} + \Xi_q\right)$ term, which is identical to those found for $\pi^\pm$, Eq. (45).